\documentclass[prd,amssymb,amsmath,amsfonts,superscriptaddress,nofootinbib,reprint,showpacs,floatfix]{revtex4-1}

\usepackage{graphicx}
\usepackage{lmodern}
\usepackage{amsmath,amssymb}
\usepackage{aas_macros}
\usepackage{mathrsfs}
\usepackage{amsfonts}
\usepackage[utf8]{inputenc}
\usepackage{url}
\usepackage[colorlinks]{hyperref}
\usepackage{xcolor,colortbl}
\usepackage{multirow}
\usepackage[normalem]{ulem}
\usepackage{rotating}
\definecolor{dodgerblue}{HTML}{1E90FF}
\definecolor{viennared}{HTML}{DA0A14}
\definecolor{ctorange}{HTML}{FF6C0C}
\definecolor{wales}{HTML}{ff0038}
\definecolor{benettongreen}{HTML}{009421}
\definecolor{ferrarired}{HTML}{ff2800}
\definecolor{austriawienpurple}{HTML}{441678}
\definecolor{gray}{HTML}{F0F0F0}
\definecolor{LightCyan}{rgb}{0.88,1,1}
\newcolumntype{a}{>{\columncolor{gray}}c}
\newcolumntype{b}{>{\columncolor{white}}c}

\hypersetup{
     colorlinks=true,
     linkcolor=dodgerblue,
     filecolor=dodgerblue,
     citecolor = dodgerblue,      
     urlcolor=dodgerblue,
     }

\newcommand{\Birmingham}{School of Physics and Astronomy and Institute for Gravitational Wave Astronomy, University of Birmingham, Edgbaston, Birmingham, B15 2TT, United Kingdom}
\newcommand{\msun}{$M_\odot$}

\begin{document}
%%%%%%%%%%

\title{Measuring precession in asymmetric compact binaries}

\author{Geraint Pratten}
\email{gpratten@star.sr.bham.ac.uk}
\affiliation{\Birmingham}

\author{Patricia Schmidt}
\email{pschmidt@star.sr.bham.ac.uk}
\affiliation{\Birmingham}

\author{Riccardo Buscicchio}
\email{riccardo@star.sr.bham.ac.uk}
\affiliation{\Birmingham}

\author{Lucy M. Thomas}
\email{lthomas@star.sr.bham.ac.uk}
\affiliation{\Birmingham}

\date{\today}

%\begin{flushright}
%LIGO-P
%\end{flushright}

%%%%%%%%%%%%%%%
\begin{abstract}
Gravitational-wave observations of merging compact binaries hold the key to precision measurements of the objects' masses and spins. General-relativistic precession, caused by spins misaligned with the orbital angular momentum, is considered a crucial tracer for determining the binary's formation history and environment, and it also improves mass estimates -- its measurement is therefore of particular interest with wide-ranging implications. 
Precession leaves a characteristic signature in the emitted gravitational-wave signal that is even more pronounced in binaries with highly unequal masses. The recent observations of GW190412 and GW190814 have confirmed the existence of such asymmetric compact binaries. 
Here, we perform a systematic study to assess the confidence in measuring precession in gravitational-wave observations of high mass ratio binaries and, our ability to measure the mass of the lighter companion in neutron star -- black hole type systems. Using Bayesian model selection, we show that precession can be decisively identified for low-mass binaries with mass ratios as low as $1:3$ and mildly precessing spins with magnitudes $\lesssim 0.4$, even in the presence of systematic waveform errors.
\end{abstract}

%%%%%%%%%%%%%%%
\maketitle 

%%%%%%%%%%%%%%% INTRODUCTION
\section{Introduction}
\label{sec:intro}
%%%%%%%%%%%%%%%
The LIGO Scientific and Virgo collaborations have recently reported the first clear detections of gravitational waves (GW) from coalescing compact binaries with unequal masses, GW190412 \cite{GW190412} and GW190814 \cite{GW190814}. The inferred mass ratio\footnote{Note that we adopt the inverse convention to that used in \cite{GW190412} and \cite{GW190814}, where $q = m_2 / m_1$.}, $q = m_1/m_2 \geq 1$, for both events points to highly asymmetric compact binary systems, differing from the binary black holes observed during the first two observing runs O1 \cite{TheLIGOScientific:2016pea} and O2 \cite{LIGOScientific:2018mvr}. Of these two events, GW190412 is consistent with a binary black hole merger, with a primary source mass of $m_1 \simeq 30 M_{\odot}$ and a secondary source mass of $m_2 \simeq 8M_{\odot}$. The second event, GW190814, is consistent with the merger of a neutron star -- black hole (NSBH) or black hole binary (BBH)~\cite{GW190814}, with a primary source mass of $m_1 \simeq 23 M_{\odot}$ and a secondary mass, $m_2 \simeq 2.6 M_{\odot}$. Notably, the mass of the secondary lies in the lower mass gap of $2.5-5 M_{\odot}$ \cite{Bailyn:1997xt,Ozel:2010apj,Farr:2011mds,Ozel:2012ax,Kreidberg2012:mmb}, making it either the heaviest neutron star (NS) or the lightest black hole (BH) observed to date~\cite{Freire:2007jd,Ozel:2010apj,Farr:2011mds,Ozel:2012ax,Kreidberg2012:mmb,Corral-Santana:2015fud,Alsing:2017bbc,Shibata:2017xdx,LIGOScientific:2019eut,Cromartie:2019kug,Essick:2019ldf,Abbott:2020uma}.
A coincident observation of an electromagnetic (EM) counterpart, such as a gamma-ray burst or a kilonova, would indicate strongly that the lighter compact object was a neutron star~\cite{Li:1998bw,Rosswog:2005su,Shibata:2006nm,Metzger:2010sy,Foucart:2012nc,Pannarale:2014rea,Arcavi:2017xiz,GBM:2017lvd,Foucart:2018rjc,Ackley:2020qkz,Kruger:2020gig}. Alternatively, we may hope to see the tidal disruption of the NS in an NSBH system \cite{Bildsten:1992my,Vallisneri:1999nq,Faber:2005yg,Shibata:2006bs,Etienne:2008re,Shibata:2009cn, Ferrari:2009bw, Shibata:2011jka,Foucart:2010eq,Kyutoku:2010zd,Kyutoku:2011vz,Foucart:2012vn,Foucart:2012nc,Foucart:2019bxj}, which would leave a characteristic imprint in the emitted GW signal. However, as the mass ratio increases, tidal effects become highly suppressed and the NS can be swallowed entirely before tidal disruption has taken place, making a highly asymmetric NSBH merger indistinguishable from a BBH merger \cite{Shibata:2011jka}. In addition, tidal effects in the early-inspiral are anticipated to be negligible for such high mass ratio binaries and we are therefore reliant on the measurement of other intrinsic parameters such as masses and spins to determine the binary composition. Accurate measurements of the component masses and spins will be important to discriminate as to whether the lighter compact object is consistent with the theoretical limits on NS masses \cite{Ozel:2012ax} or with a low-mass BH \cite{Farr:2011mds}.

While the formation of compact binaries with asymmetric mass ratios is highly uncertain, such unequal-mass binaries are of particular interest for measuring relativistic spin effects such as spin-precession. 
Spin-precession is sourced by the misalignment between the orbital angular momentum of the binary motion and the individual spins of the two compact objects, inducing additional structure in the form of characteristic modulations in the GW signal~\cite{Apostolatos:1994, Kidder:1995zr}. This helps breaking correlations such as the mass -- spin degeneracy, in which one can mimic the effect of spin by modifying the mass ratio of the compact binary. By breaking these degeneracies, we can infer tighter mass constraints \cite{Vecchio:2003tn,Lang:1900bz,Klein:2009gza,Chatziioannou:2014coa,Vitale:2014mka,OShaughnessy:2014shr}. Moreover, the orientation of the spin angular momenta is considered one of the main tracers of a binary's formation channel and may help discerning the nature of the compact objects \cite{Kalogera:1999tq,Mandel:2009nx,Rodriguez:2015oxa,Rodriguez:2016vmx,Farr:2017uvj,Farr:2017gtv,Fishbach:2017dwv,Belczynski:2017gds,Gerosa:2017kvu,Talbot:2017yur,Zhu:2017znf,Stevenson:2017dlk,Wysocki:2018mpo,LIGOScientific:2018jsj,Kimball:2020opk}. 
GW observations so far, however, have not yielded a confident measurement of precession effects, with the events observed to date being consistent with either small spins, or large misaligned spins~\cite{LIGOScientific:2018mvr, GW190412, GW190814}. GW190814 provided the tightest constraint on precession from all GW observations to date, and constrained precession to be near-zero \cite{GW190814}.

In this paper, we reassess the confidence to which we can measure spin-precession effects in high mass ratio binaries similar to GW190814, and the confidence to which we can constrain the mass of the lighter companion at current detector sensitivity. Firstly, we investigate the degree to which we can constrain precession in GW observations of asymmetric binaries using Bayesian model selection and several statistical measures. 
Secondly, we demonstrate the efficacy of spin-precession in breaking the mass -- spin degeneracy in order to improve constraints on the mass of the secondary companion \cite{Vecchio:2003tn,Lang:1900bz,Klein:2009gza,Chatziioannou:2014coa}.
We use simulated GW signals from moderately inclined binaries with different mass ratios and varying amount of precession to study the measurability of precession effects in a systematic way. We demonstrate that even small amounts of precession can be identified confidently despite the presence of systematic errors. 
Further, using a second set of simulated signals with the smaller object between $1.25$ and $3M_\odot$ we find that the mass of the secondary is consistently underestimated for the binaries considered when spin-precession is neglected, leading to an increased risk of misidentifying a low-mass BH as a NS. 
Similar questions focusing on large populations have been addressed in previous studies~\cite{Littenberg:2015tpa, Pankow:2016udj, Chen:2019aiw}. Here, we focus primarily on GW190814-like binaries, adopting masses, inclinations and signal-to-noise ratios (SNR) broadly consistent with the values reported in \cite{GW190814}. We note that the results presented in Sec.~\ref{sec:prec} are also broadly applicable to binaries similar to GW190412. 

This paper is organised as follows: In Sec.~\ref{sec:bin} we provide an introduction to asymmetric binaries and precession. We then introduce our methodology in Sec.~\ref{sec:methods}, before presenting our main results in Sec.~\ref{sec:results}. We conclude in Sec.~\ref{sec:discussion}. 
Throughout this paper we set $G=c=1$ unless stated otherwise.

%%%%%%%%%%%%%%% GW190814-like binaries
\section{Asymmetric compact binaries}
\label{sec:bin}
%%%%%%%%%%%%%%%
The current understanding of binary evolution leads to a number of distinct binary formation channels. Proposed scenarios include isolated \cite{Dominik:2014yma,Belczynski:2016obo,Eldridge:2016iso,Stevenson:2017tfq}, dynamical \cite{PortegiesZwart:2002iks,Gultekin:2005fd,Mapelli:2016vca,Bartos:2016dgn,Rodriguez:2016avt,Chatterjee:2016thb,Zevin:2018kzq,Sedda:2020jvg,McKernan:2020lgr}, and primordial \cite{Bird:2016dcv,Clesse:2016vqa} formation with many sub-channels within each category. Each of these formation channels will leave a characteristic imprint on the mass \cite{TheLIGOScientific:2016htt,Kovetz:2016kpi,Fishbach:2017zga,Wysocki:2018mpo,Talbot:2018cva}, spin \cite{Mandel:2009nx,Stevenson:2017dlk,Farr:2017uvj,Talbot:2017yur} and redshift \cite{Sathyaprakash:2009xt,Sathyaprakash:2012jk,Taylor:2012db,Dominik:2013tma,Rodriguez:2018rmd,Fishbach:2018edt,Vitale:2018yhm} distributions of the observed compact binaries. 

Modelling of the mass distribution using the ten BBHs detected during the first two observing runs \cite{LIGOScientific:2018mvr} finds a median mass ratio of $q = 1.1$ at $90\%$ credibility and predicts that $99\%$ of binaries detected will have mass ratios $q < 2$ \cite{Fishbach:2019bbm}. This makes the recent observation of GW190814, a highly asymmetric binary with a mass ratio of $q \sim 9$, something of an enigma. Plausible formation channels for such asymmetric binaries include dynamical \cite{DiCarlo:2020lfa,Hamers:2020huo} and hierarchical merger scenarios \cite{Fishbach:2017dwv,Gerosa:2017kvu,Rodriguez:2019huv,Gerosa:2020bjb,McKernan:2020lgr}. For isolated binary formation channels, the prevalence of asymmetric compact binaries can be sensitive to the metallicity of the environment, with asymmetric binaries being preferred in low metallicity environments \cite{Dominik:2012kk,Stevenson:2017tfq,Giacobbo:2017qhh}. Accretion disks of active galactic nuclei (AGN) could be promising environments for driving hierarchical mergers, in which asymmetric binaries are likely \cite{Yang:2019cbr}. 

Furthermore, while the primary mass allows us to identify the heavier component as a black hole, the secondary mass is compatible with being either a BH or a NS. We note, however, that the lighter companion with $\sim 2.6 M_\odot$ is at the threshold of the maximum theoretically supported NS mass~\cite{Lattimer:1990apj,Chamel:2013efa,Rezzolla:2017aly,LIGOScientific:2019eut} and is in tension with current constraints from the maximum NS masses inferred from GW170817 and pulsar observations \cite{Freire:2007jd,LIGOScientific:2019eut,Cromartie:2019kug,Essick:2019ldf}. In addition, the mass of the secondary is comparable to the BH masses
created as binary neutron star merger products \cite{TheLIGOScientific:2017qsa,LIGOScientific:2019eut,Gupta:2019nwj} as well as a recently reported low-mass BH in a non-interacting BH-giant star binary \cite{Thompson:2018ycv}. Population synthesis models for the formation of NSBH binaries demonstrate a preference for a system comprising a heavy NS ($m_{\rm NS} \sim 1.3 - 2.0M_{\odot}$) and a low mass BH ($m_{\rm BH} \sim 5 - 15 M_{\odot}$), especially for formation channels with low natal kicks \cite{Giacobbo:2018etu}. Such binaries would correspond to mass ratios $q \sim 3 - 8$, further emphasising the need to understand the confidence to which we can infer the intrinsic properties of asymmetric binaries from GW observations. 

Another source of asymmetry, besides unequal masses, pertains to the spins of the two companions, which are of particular interest for discriminating between different formation channels. 
Binaries that form through dynamical interactions are anticipated to have isotropically oriented spins. This is in stark contrast to binaries that form from isolated compact objects, where spins are preferentially aligned with the orbital angular momentum. For isolated binaries, supernova kicks are one of the primary mechanism that give rise to spins misaligned with the orbital angular momenta \cite{Kalogera:1999tq}. Constraints on precession in compact binaries can therefore significantly shape our understanding of binary formation channels and their evolution.

In order to infer the source properties from GW observations, highly accurate waveform models that govern the the inspiral, merger and ringdown are necessary. The GW signals of compact binaries with highly asymmetric masses possesses a rich phenomenology due to the excitation of higher-order multipoles.
The higher-order modes of the gravitational field encode additional information about the source which allows for the breaking of certain parameter degeneracies, such as the inclination -- distance correlation~\cite{Markovi:1993inc,Cutler:1994fla,Nissanke:2009kt}. 

Binaries whose spins are aligned with the orbital angular momentum, exhibit strong correlations between the masses/mass ratio and spins~\cite{Baird:2012cu}. Arbitrarily oriented spins, however, break the equatorial symmetry of the binary system and induce general relativistic spin-precession~\cite{Barker:1975ae, Thorne:1984mz, Apostolatos:1994, Kidder:1995zr}. These affect the emitted signal in several ways: (i) they leave characteristic imprints in the form of amplitude and phase modulations; (ii) they modify the final state of the remnant; (iii) they excite higher-order modes. Similar to unequal masses, precession of the orbital plane allows us to break another parameter correlation, the mass -- spin degeneracy~\cite{Vecchio:2003tn,Lang:1900bz,Chatziioannou:2014coa}. Spin precession could therefore be of particular importance when one seeks to distinguish between NSs and low-mass BHs in the absence of a clear tidal signature.

In what follows, we will be considering simulated signals that contain both higher-order modes and precession in order to mimick a realistic scenario as best as possible. 

%%%%%%%%%%%%%%% METHODS
\section{Methodology}
\label{sec:methods}
%%%%%%%%%%%%%%% PRECESSION
\subsection{Effective Precession Spin}
\label{sec:prec}
%%%%%%%%%%%%%%%
Coalescing BBHs on quasi-spherical orbits are intrinsically characterised by their mass ratio $q = m_1/m_2 \geq 1$, where $m_i$ is the component mass of the i-th black hole, and their (dimensionless) spin angular momenta $\vec{\chi}_i$. 
The dominant spin effect on the inspiral rate is captured by the effective aligned spin, a mass-weighted combination of the spins parallel to the orbital angular momentum $\hat{L}$ \cite{Racine:2008qv,Ajith:2009bn,Santamaria:2010yb}
\begin{align}
    \chi_{\rm eff} &= \frac{m_1 \chi_1 + m_2 \chi_2}{m_1 + m_2} .
\end{align}
\newline
Depending on the binary's formation history, however, the spins may be arbitrarily oriented with respect to the orbital angular momentum $\hat{L}$~\cite{Farr:2017uvj}. Misalignment between the spins and $\hat{L}$ induces general relativistic precession of the orbital plane and spins~\cite{Apostolatos:1994,Kidder:1995zr}, i.e. the (four) spin components perpendicular to $\hat{L}$ source these precession effects. Over many GW cycles, these spin components contained within the instantaneous orbital plane may be approximated by a scalar quantity, $\chi_p$, which captures the average amount of precession in a binary system~\cite{Schmidt:2014iyl} defined as:
\begin{equation}
\label{eq:chip}
    \chi_p := \frac{1}{A_1 m_1^2}\max(A_1 S_{1\perp}, A_2 S_{2\perp}),
\end{equation}
where $\vec{S}_i = m_i^2 \vec{\chi}_i$, $S_{i\perp} = || \hat{L} \times (\vec{S}_i \times \hat{L})||$, $A_1 = 2+3q/2 $ and $A_2 = 2+3/(2q)$. The effective precession spin $\chi_p$ is defined in the domain $[0,1]$, where $\chi_p =0$ corresponds to a non-precessing and $\chi_p=1$ to a maximally precessing binary. It is important to note, however, that even very strongly precessing binaries may not be easily identified as such if the line of sight is approximately along the direction of the total angular momentum, as imprint of precession on the GW signal will be minimized \cite{Schmidt:2012rh}. 
We will use the effective precession spin $\chi_p$ in our analyses to characterise the amount of precession present in a binary system, and statements concerning the measurability of precession will be based on its inferred distribution.

%%%%%%%%%%%%%%% PRECESSING SNR
\subsection{Precessing SNR}
\label{sec:rhop}
%%%%%%%%%%%%%%%
The strength of an observed GW signal $h$ is characterised by its signal-to-noise ratio (SNR) $\rho$ defined as:
\begin{equation}
    \label{eq:SNR}
    \rho := \sqrt{\langle h|h\rangle} = 2 \Bigg[ \int_0^\infty df \frac{|\tilde{h} (f)|^2}{S_n(f)}\Bigg]^{1/2}, 
\end{equation}
where $\tilde{h} (f)$ denotes the Fourier transform of $h$ and $S_n(f)$ is the noise power spectral density (PSD). 

Recently, Ref.~\cite{Fairhurst:2019har} introduced a frequentist framework to estimate the contribution to the SNR $\rho$ that stems from precession, referred to as precessing SNR $\rho_p$. The formalism decomposes the GW signal into two harmonics, each of which is equivalent to the emission of a non-precessing binary. The modulations typical for a precessing system are introduced through the beating between the two harmonics. $\rho_p$ is then defined as the SNR contained in the harmonic orthogonal to the dominant one. In the absence of precession, $\rho_p$ is $\chi^2$-distributed with two degrees of freedom. 
A simple criterion for precession to be considered observable is the requirement that $\rho_p \geq 2.1$~\cite{Fairhurst:2019har, Fairhurst:2019srr}. Here, we will assess the significance of $\rho_p$ via the single-sided $p$-value associated with the mean of the distribution.

The two harmonics formalism relies on several assumptions ~\cite{Fairhurst:2019har}, which are valid for the signals considered in this paper.
We will thus use it as a complementary quantifier to assess the measurability of spin precession. We stress, however, that $\rho_p$ is an inherently frequentist quantity, while our main analyses will be fully Bayesian as discussed in Sec.~\ref{sec:bayes}.

%%%%%%%%%%%%%%% INJECTIONS
\subsection{Simulated Gravitational-Wave Signals}
\label{sec:inj}
%%%%%%%%%%%%%%%
\begin{table}[t!]
    \centering
    \begin{tabular}{l|c}
    \hline
    \hline
         Parameter & Value  \\
         \hline
         Chirp mass $\mathcal{M}_c$ [\msun] & 6.3 \\
         Effective inspiral spin $\chi_{\rm eff} $ & 0.0 \\
         Inclination $\iota$ [rad] & 0.70 \\
         RA $\alpha$ [rad] & 0.23 \\
         DEC $\delta$ [rad] & -0.42 \\
         Polarisation $\psi$ [rad] & 3.0 \\
         \hline
         SNR $\rho$ & 30 \\
         \hline
         \hline 
    \end{tabular}
    \caption{Fixed parameter values for all simulated GW signals in the first data set. The mass -- spin degeneracy data set has the same extrinsic parameters, SNR and $\chi_{\rm eff}$ but fixes $\chi_p=0.2$ and varies $\mathcal{M}_c$ instead (see main text). These values are consistent with the parameters inferred for GW190814~\cite{GW190814,gw190814-data}.}
    \label{tab:injparams}
\end{table}

We create two sets of simulated GW signals (injections),  which include both precession and a subset of higher-order modes as expected for real signals. We inject the signals into zero-noise, which is representative of the results when averaging over identical injections in different Gaussian noise realizations. All mock signals used in our analyses are generated from the effective-one-body (EOB) waveform model \texttt{SEOBNRv4PHM}~\cite{Ossokine:2020v4phm} for binary black holes\footnote{This waveform model does not contain tidal effects, which are negligible for the high mass ratios considered in our analysis. Tidal disruption could in principle occur for some of the lower mass ratio binaries but is not taken into account here.}. The EOB framework~\cite{Buonanno:1998gg,Buonanno:2000ef,Damour:2000we,Damour:2001tu} models the complete inspiral-merger-ringdown GW signal of coalescing compact binaries in the time-domain. It utilises analytical information from post-Newtonian theory and gravitational self force and is tuned to numerical relativity (NR) in the strong field regime. We note that neither the EOB model nor the recovery waveforms described in Sec.~\ref{sec:bayes} are calibrated against precessing NR simulations. Precessing NR simulations at high mass ratios are numerically challenging leading to a lack of waveforms in this region of the binary parameter space. We therefore use \texttt{SEOBNRv4PHM} as our injection model as it incorporates full spin degrees of freedom, higher-order modes and is demonstrably robust at high mass ratios, which is crucial for our study.

The first set of injections has a varying mass ratio $q \in [3,10]$ and $\chi_p \in [0.0, 0.4]$ chosen such that the lighter companion is always non-spinning. 
All other parameters are fixed and listed in Tab.~\ref{tab:injparams}.
They are chosen to be consistent with GW190814 \cite{GW190814}, in particular the source frame chirp mass, $\mathcal{M}_c = (m_1 m_2)^{3/5}/(m_1+m_2)^{1/5}$, and the inclination. 
Consistent with the majority of observed signals to date~\cite{LIGOScientific:2018mvr}, we only consider binaries with a vanishing inspiral spin $\chi_{\rm eff}$~\cite{Qin:2018vaa,Fuller:2019sxi,Miller:2020zox}. We do not expect this particular choice to affect our results due to the approximate decoupling between the inspiral and precession dynamics~\cite{Schmidt:2010it, Schmidt:2012rh}.

Furthermore, our injections have a fixed SNR of $\rho=30$, representing moderately loud signals at current and near-future detector sensitivities \cite{Aasi:2013wya}. 
For mass ratio $q=9$ we create additional injections with $\rho=10$ and $\rho=20$; since we fix the binary inclination to a moderate value of $\sim 40^\circ$, this amounts to changing the luminosity distance to adjust the SNR. 

A second set of injections explores the mass -- spin degeneracy briefly discussed in Sec.~\ref{sec:bin}. Here we pin the mass of the primary to $m_{1} = 20 M_{\odot}$ and vary the mass of the secondary in the range $m_2 \in \left[ 1.25, 1.5, \dots , 3 \right] M_{\odot}$. The effective inspiral spin is fixed at $\chi_{\rm eff} = 0$ and we allow for small non-vanishing spin-precession with $\chi_p = 0.2$.  All other parameters are identical to the first set. This series is chosen to span a range of astrophysically interesting component masses that graze the lower boundary of the mass gap,
and serves to highlight the importance of precessing waveform models in constraining the component masses, especially near the maximum theoretical NS mass.

%%%%%%%%%%%%%%% MODEL SELECTION
\subsection{Bayesian Inference \& Model Selection}
\label{sec:bayes}
%%%%%%%%%%%%%%% 
We treat the measurability of precession in an asymmetric binary system as a Bayesian model selection problem. The probability of obtaining the binary parameters $\theta$ given the data $d$ and a signal model hypothesis $\mathcal{H}$ is
\begin{align}
    p\left(\theta | d, \mathcal{H} \right) = \frac{\mathcal{L} \left( d | \theta, \mathcal{H} \right) \pi \left(\theta | \mathcal{H} \right)}{\mathcal{Z_{\mathcal{H}}}},
\end{align}
\newline 
where $\mathcal{L}$ is the likelihood, $\pi$ the prior and $\mathcal{Z}$ the signal evidence
\begin{align}
    \mathcal{Z}_{\mathcal{H}} \equiv \int d \theta \mathcal{L}(d | \mathcal{H}, \theta) \pi(\theta | \mathcal{H}) ,
\end{align}
such that the noise evidence $\mathcal{Z}_n$ is defined by
\begin{align}
\mathcal{Z}_{n} &\equiv \mathcal{L}(d | n ) .
\end{align}
The Bayes factor $\mathcal{B}$ for a signal, assuming a model hypothesis $\mathcal{H}$, over noise $n$ is
\begin{align}
\label{eq:BF}
    \mathcal{B}_{\mathcal{H}/n} &= \frac{\mathcal{Z}_{\mathcal{H} }}{\mathcal{Z}_n} .
\end{align}
\newline 
In this analysis, we will be interested in comparing the evidence for the precessing hypothesis $\mathcal{H} = p$ against the non-precessing hypothesis $\mathcal{H} = np$,
\begin{align}
\label{eq:precBayes}
    \mathcal{B}_{p/np} &= \frac{\mathcal{Z}_{p}}{\mathcal{Z}_{n}} \,  \frac{\mathcal{Z}_{n}}{\mathcal{Z}_{np}} = \frac{\mathcal{Z}_{p}}{\mathcal{Z}_{np}}.
\end{align}

%%%%
%FIGURE DUMP
\begin{figure*}[th!]
\includegraphics[width=2\columnwidth]{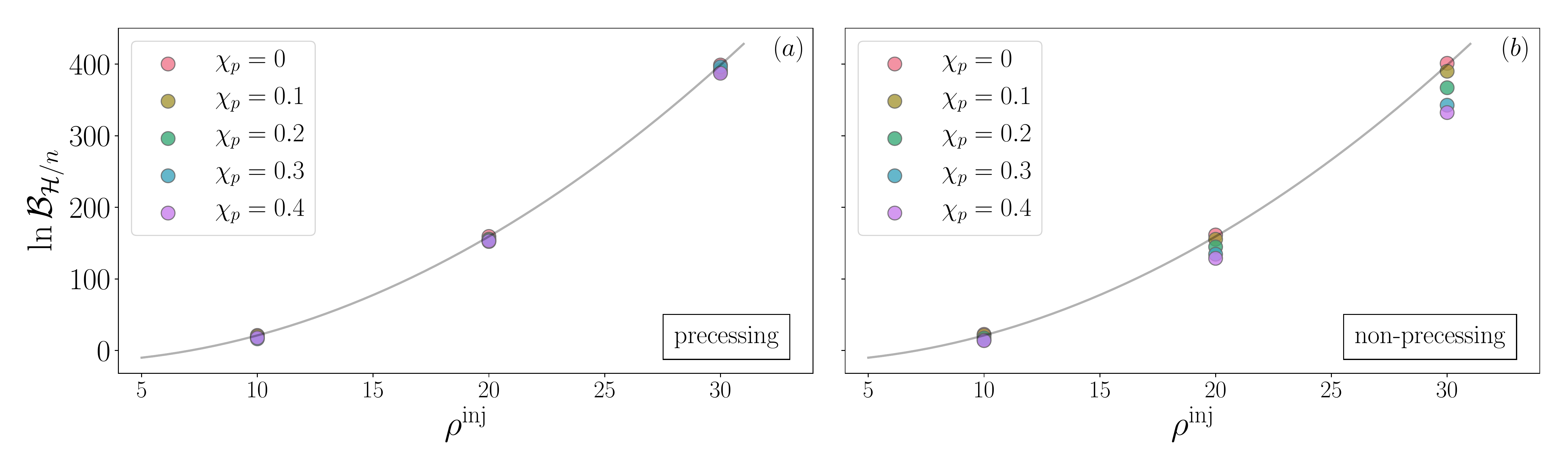}
\caption{Signal vs. noise Bayes factor as a function of injected SNR $\rho^{\rm inj}$ for \texttt{PhenomPv2} (a) and \texttt{PhenomD} (b) for $q=9$ binaries for different values of $\chi_p$. The expected scaling $\ln \mathcal{B}_{\mathcal{H}/n} \propto (\rho^{\rm inj})^2$ (grey solid curve)
% expected quadratic dependence on the SNR 
is only recovered for the precessing waveform model (a), with significant deviations observed for the aligned-spin waveform model for $\chi_p \geq 0.2$ and $\rho > 10$ (b) due to the neglect of precession.}
\label{fig:q9BF}
\end{figure*}
%%%%

%%% models
We perform Bayesian inference~\cite{Bayes:1764vd} on our simulated signals using the nested sampling algorithm \cite{Skilling:2006gxv,VeitchVecchio:2008prd,VeitchVecchio:2008prd,VeitchVecchio:2010} implemented in the publicly available inference library \texttt{LALInference}~\cite{Veitch:2014wba}. We inject the simulated signals into a zero-noise LIGO-Virgo three-detector network with a sensitivity representative of the first three months of the third observing run~\cite{Tse:2019wcy,Acernese:2019sbr,NoiseCurves}. We marginalise over calibration uncertainties \cite{Vitale:2011wu,Cahillane:2017vkb,Sun:2020wke} using the representative values reported in \cite{LIGOScientific:2018mvr}, and start the likelihood integration at 20Hz. Our signal hypotheses will be two phenomenological waveform models, \texttt{IMRPhenomD} (non-precessing)~\cite{Khan:2015jqa, Husa:2015iqa} and \texttt{IMRPhenomPv2} (precessing)~\cite{Hannam:2013oca}. We note that these two waveform models are not independent of each other; \texttt{IMRPhenomPv2} is obtained by applying a rotation transformation to the quadrupolar modes of \texttt{IMRPhenomD} following the framework developed in Refs.~\cite{Schmidt:2010it, Schmidt:2012rh, Schmidt:2014iyl}.
The two phenomenological waveform models, however, differ in various aspects from our simulated signals, for example they do not include higher-order modes and \texttt{IMRPhenomPv2} uses fewer spin degrees of freedom to model precession, 
hence systematic modelling errors due to inaccurate modelling or neglected physics are included in our analyses. 

For the priors, we follow the choices as detailed in App. B of \cite{LIGOScientific:2018mvr}. We use uniform priors on the component masses $m_i \in [1,40]M_{\odot}$, isotropic priors on the spin orientations and a uniform prior on the dimensionless spin magnitudes $\chi_i \leq 0.99$\footnote{We note that this includes binaries with $\chi_2 > 0.7$, which is larger than the maximally allowed spin for neutron stars. However, we consider this choice appropriate due to the unknown nature of the secondary object.}. To enable a direct comparison to the precessing approximant, we use the z-prior for the spin priors, e.g. App.~A of \cite{Lange:2018pyp}, for \texttt{IMRPhenomD}. For the distance, we adopt a prior proportional to the luminosity distance squared with an upper cutoff of $600 \rm{Mpc}$. 

%%% bias
From the one-dimensional posterior probability distribution function (PDF) we can obtain the parameter biases induced by systematics. Specifically, we define the bias as the difference between the maximum a posteriori (map) value of a parameter $x$ and its true value, i.e.,
\begin{equation}
    \Delta x := x^{\rm map} - x^{\rm true}.
\end{equation}
For the effective precession spin parameter $\chi_p$ it follows that if $\Delta \chi_p > 0$ the amount of precession in the system is overestimated and if $\Delta \chi_p < 0$ it is underestimated.

%%% posterior quantile
Additionally, we also use the posterior quantile of the true parameter value $x^\mathrm{true}$ given by
\begin{equation}
\label{eq:quantile}
    Q(p) :=\frac{1}{2} - \int_{x^\mathrm{min}}^{x^\mathrm{true}}  p(x | d,\mathcal{H}) \in [-0.5,0.5],
\end{equation}
as a measure of the displacement between the posterior median and the true value. For precession-related parameters $Q>0$ ($Q<0$) implies an overestimation (understimation) of the amount of precession in the binary system. Moreover, the quantile also encodes the skew of the distribution.

%%% Jensen-Shannon divergence
To ascertain confidence in the measurement of precession, we additionally employ two statistical measures: the Kullback-Leibler divergence ($D^x_{\rm KL}$)~\cite{Kullback:1951} and the (related) Jensen-Shannon divergence ($D^x_{\rm JS}$)~\cite{Lin91divergencemeasures}.
These two measures allow us to quantify the difference between two probability distribution $p(x)$ and $q(x)$ and are used to measure the information gain between the prior and the posterior distribution of a continuous random variable $x$. The Kullback-Leibler divergence is defined as, 
\begin{equation}
\label{eq:DKL}
    D^x_{\rm KL}(p | q) = \int p(x) \log_2 \Bigg(\frac{p(x)}{q(x)}\Bigg) \mathrm{d}x.
\end{equation}
The Jensen-Shannon divergence, which defines a natural, normalised distance measure between two distributions, is given by
\begin{equation}
\label{eq:DJS}
    D^x_{\rm JS} (p | q) = \frac{1}{2}\Bigg( D^x_{\rm KL}(p|s) + D^x_{\rm KL}(q|s)\Bigg),
\end{equation}
where $s = (1/2)(p+q)$.

%%%%%%%%%%%%%%% RESULTS
\section{Results}
\label{sec:results}
%%%%%%%%%%%%%%%

%FIGURE DUMP
\begin{figure*}[ht!]
\includegraphics[width=\textwidth]{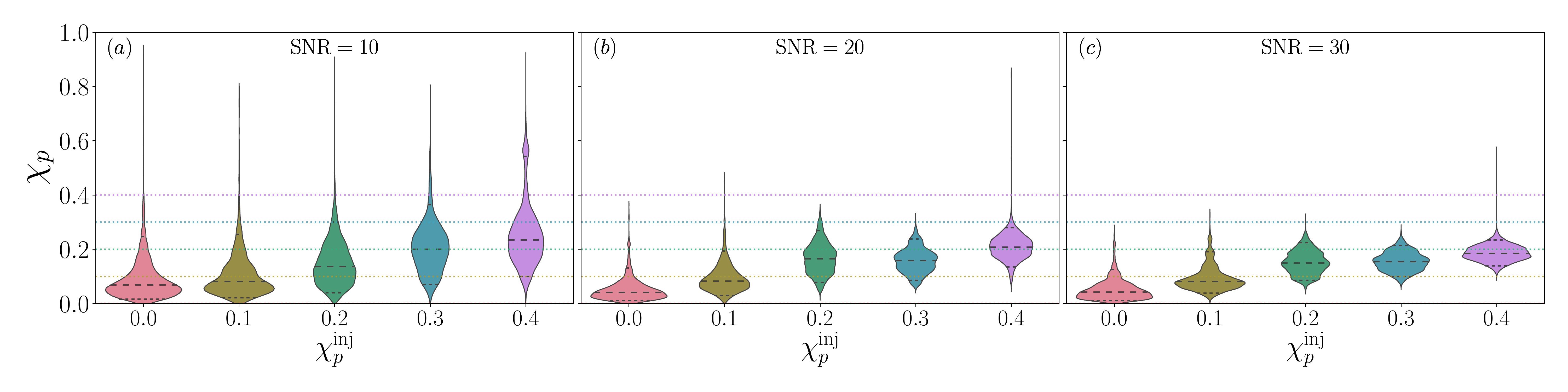}
\caption{One-dimensional PDF of the precession parameter $\chi_p$ for binaries with $q=9$ and varying amounts of precession (indicated on the $x$-axis) at three SNRs: (a) $\rho=10$, (b) $\rho=20$ and (c) $\rho=30$. The horizontal lines indicate the median and 90\% CI. We see that the posterior width decreases as the SNR increases while the median changes only slightly, showing a clear systematic difference between the true $\chi_p$-value and the inferred median.}
\label{fig:q9SNR}
\end{figure*}

%%%%%%%%%%%%%%%%
\subsection{Precession Measurements}
\label{sec:prec}
%%%%%%%%%%%%%%%%
The Bayes factor between two hypotheses is a commonly used discriminator to assign confidence to a particular hypothesis, e.g. \cite{Kass:1995loi,gelmanbda04}. Here, we treat the measurability of precession as a Bayesian model selection problem and use the Bayes factor between the precessing and the non-precessing hypotheses to quantify the confidence to which we can measure precession. We first examine in detail a binary system with mass ratio $q=9$, consistent with the inferred mass ratio of GW190814~\cite{GW190814}. In particular, we investigate the measurability of precession as a function of injected SNR and precession spin $\chi_p$.

Figure~\ref{fig:q9BF} shows the signal versus noise Bayes factor (Eq.\eqref{eq:BF}) as a function of the injected signal SNR $\rho^{\rm inj}$ for the precessing and the non-precessing recovery models for different values of $\chi_p$. The Bayes factor for the signal to noise hypothesis approximately scales as $\ln \mathcal{B}_{\mathcal{H}/n} \propto (\rho^{\rm inj})^2$, see e.g. \cite{Cornish:2011ys}. For all values of $\chi_p$, we observe such a scaling when using the precessing waveform model. The non-precessing model, however, shows significant deviations from this relation, especially for larger values of $\chi_p$ and with increasing SNR, where the non-precessing waveform model systematically underestimates the injected SNR due to missing physics in the waveform approximant. In particular, we recall that neither recovery waveform model includes higher-order modes, while our simulated signals do. The results in Fig.~\ref{fig:q9BF} suggest, however, that higher-order modes play a subdominant role in comparison to precession (see also \cite{Cho:2012ed}).
As a point of caution, we note that at high mass ratios and high $\chi_p$, \texttt{IMRPhenomD} shows strong systematic biases towards higher mass ratios. If we do not take appropriate care when choosing our priors, e.g. for $q$, this can lead to significant railing that impacts the calculation of the Bayes factors and the results can become unreliable.

In Fig.~\ref{fig:q9SNR} we show the one-dimensional posterior distributions for the effective spin parameter $\chi_p$ for the $q=9$ series at three different injected SNRs. We find that the precessing waveform model \texttt{IMRPhenomPv2} systematically underestimates $\chi_p$ except for the non-precessing case, i.e. $\chi_p = 0$. We note, however, that this may be different for other binary inclinations. 

As expected, with increasing SNR, tighter 90\% credible interval (CI) bounds are obtained, and we find that the posterior widths scale $\propto \rho^{-1}$, as anticipated in the high-SNR limit \cite{Cutler:1994fla,Poisson:1995ef}. The result in Fig.~\ref{fig:q9SNR} also indicates that at higher SNRs, we can more confidently exclude the non-precessing limit for smaller values of $\chi_p$ due to the reduction in posterior support as $\chi_p \rightarrow 0$. We note that the results discussed here correspond to injections in zero-noise, as would be expected if averaged over many noise-realizations. Individual noise realizations will induce a spread in the Bayes factors, though we leave a detailed characterization to further study. 

\begin{figure}[t!]
\includegraphics[width=\columnwidth]{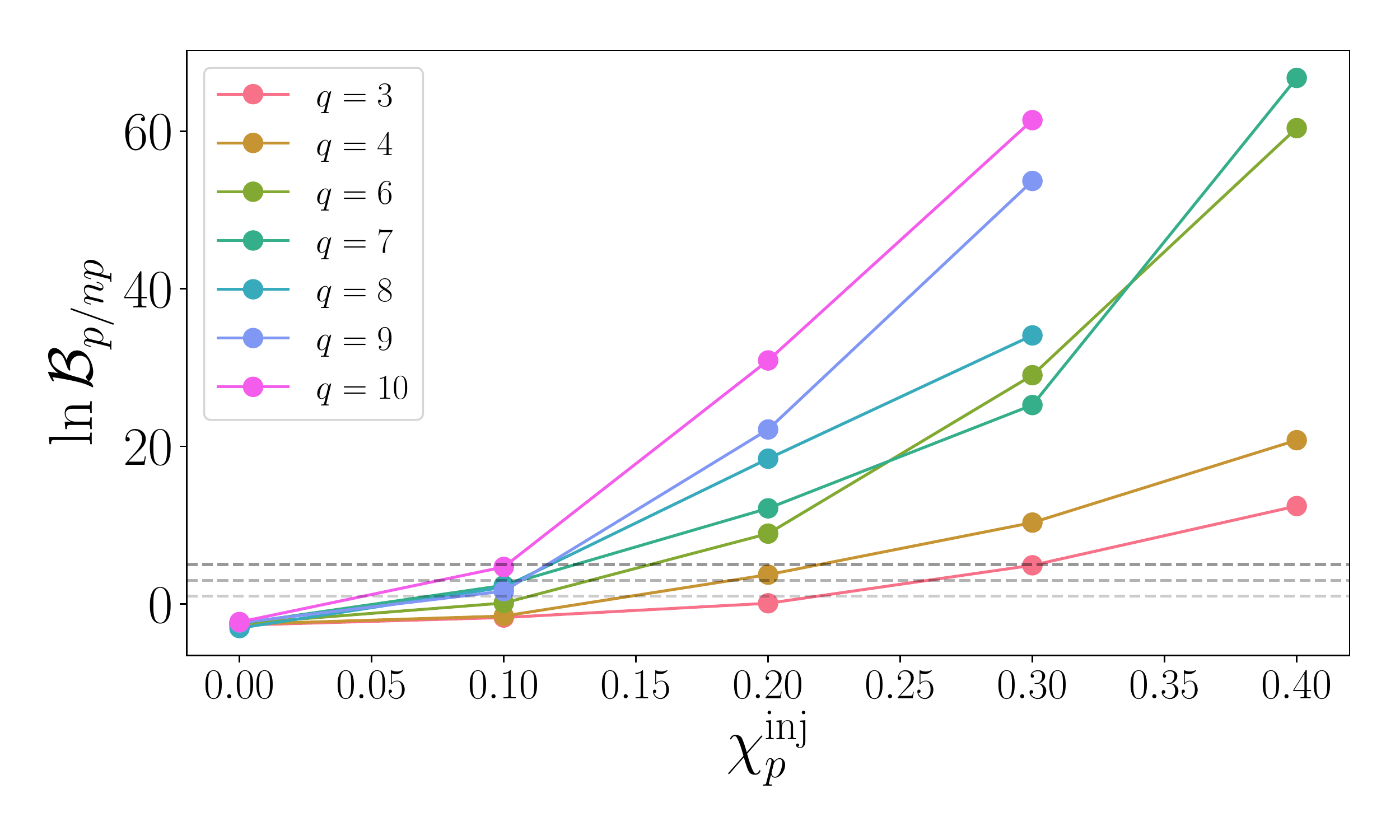}
\caption{Bayes factor for the precessing waveform model against the non-precessing waveform model for varying mass ratio $q$ and the injected effective precession spin parameter $\chi_p$. Following Ref.~\cite{Kass:1995loi}, the horizontal dashed lines indicate positive ($1.0$), strong ($3.0$) and very strong ($5.0$) evidence against the non-precessing hypothesis.}
\label{fig:fLBayes}
\end{figure}

We now turn to the analysis of the larger ensemble of mock GW signals described in Sec.~\ref{sec:inj}. We recall that we have fixed the SNR to $\rho^{\rm inj}=30$, which corresponds to moderately loud signals for current and near-future detector sensitivities~\cite{Aasi:2013wya}.
Figure~\ref{fig:fLBayes} shows the Bayes factor for the precessing versus the non-precessing signal hypothesis, Eq.~\eqref{eq:precBayes}, as a function of $\chi_p$ for all mass ratios considered. We find that for $q > 5$ the precessing signal hypothesis is strongly favoured (i.e. $\ln \mathcal{B}_{p/np} \geq 5$) for $\chi_p \geq 0.2$. For lower mass ratios, a larger amount of precession, i.e. $\chi_p > 0.3$, is required to clearly differentiate the non-precessing from the precessing hypothesis. 
Similar to our observation for the $q=9$ case, we find that the Bayes factor becomes unreliable for high mass ratios and large amounts of precession, hence the data points for $\chi_p = 0.4$ and $q \geq 8$ are omitted.

% Table with various statistics parameters for \chi_p
\begin{table*}[h!t]
\setlength{\tabcolsep}{3.pt}
\centering
{\renewcommand{\arraystretch}{1.6}
\begin{tabular}{c|a b a b a|b a b a b|a b a b}
\hline
\hline
& \multicolumn{5}{ c }{$\chi_p$} & \multicolumn{5}{ |c }{$\Delta\chi_p$} & \multicolumn{4}{ |c }{$Q( \chi_p)$}\\[0.15cm]\hline
    $\chi_p^{\rm inj}$ & $0.0$ & $0.1$ & $0.2$ & $0.3$ & $0.4$ & $0.0$ & $0.1$ & $0.2$ & $0.3$ & $0.4$ & $0.1$ & $0.2$ & $0.3$ & $0.4$ \\[0.05cm]\hline
\hline
     $q=3$ & $0.12^{+0.29}_{-0.10}$ & $0.12^{+0.19}_{-0.08}$ & $0.19^{+0.22}_{-0.13}$ & $0.24^{+0.09}_{-0.10}$ & $0.28^{+0.21}_{-0.12}$ & 0.21 & 0.09 & 0.03 & -0.14 & -0.17 & 0.1 & -0.03 & -0.36 & -0.33 \\[0.07cm]
\hline
     $q=4$ & $0.1^{+0.16}_{-0.07}$ & $0.12^{+0.18}_{-0.08}$ & $0.15^{+0.13}_{-0.08}$ & $0.22^{+0.12}_{-0.08}$ & $0.2^{+0.12}_{-0.05}$ & 0.09 & 0.01 & -0.07 & -0.10 & -0.19 & 0.1 & -0.26 & -0.39 & -0.5 \\[0.07cm]
\hline
     $q=6$ & $0.07^{+0.09}_{-0.05}$ & $0.09^{+0.1}_{-0.05}$ & $0.13^{+0.07}_{-0.04}$ & $0.19^{+0.08}_{-0.07}$ & $0.22^{+0.11}_{-0.07}$ & 0.03 & 0.06 & -0.07 & -0.15 & -0.20 & -0.09 & -0.44 & -0.49 & -0.5 \\[0.07cm]
\hline 
     $q=7$ & $0.04^{+0.09}_{-0.03}$ & $0.09^{+0.08}_{-0.05}$ & $0.13^{+0.08}_{-0.05}$ & $0.18^{+0.07}_{-0.07}$ & $0.24^{+0.07}_{-0.09}$ & 0.09 & -0.01 & -0.09 & -0.15 & -0.17 & -0.07 & -0.42 & -0.49 & -0.5 \\[0.07cm]
\hline 
     $q=8$ & $0.06^{+0.08}_{-0.05}$ & $0.09^{+0.08}_{-0.04}$ & $0.11^{+0.07}_{-0.04}$ & $0.17^{+0.06}_{-0.05}$ & $0.2^{+0.05}_{-0.05}$ & 0.10 & $< -10^{-3}$& -0.03 & -0.11 & -0.2 & -0.18 & -0.48 & -0.5 & -0.5 \\[0.07cm]
\hline 
     $q=9$ & $0.04^{+0.08}_{-0.03}$ & $0.08^{+0.11}_{-0.04}$ & $0.15^{+0.08}_{-0.06}$ & $0.15^{+0.06}_{-0.06}$ & $0.18^{+0.05}_{-0.05}$ & 0.02 & -0.01 & -0.02 & -0.17 & -0.18 & -0.19 & -0.36 & -0.5 & -0.5 \\[0.07cm]
\hline
     $q=10$ & $0.03^{+0.06}_{-0.02}$ & $0.08^{+0.09}_{-0.03}$ & $0.15^{+0.07}_{-0.06}$ & $0.16^{+0.04}_{-0.05}$ & $0.18^{+0.05}_{-0.04}$ & 0.02 & 0.01 & -0.03 & -0.12 & -0.23 & -0.22 & -0.38 & -0.5 & -0.5 \\[0.07cm]
\hline
\hline
\end{tabular}
}
\caption{Median and 90\% CI, bias and quantile for $\chi_p$ for all mass ratios and injected $\chi_p$-value. We find that precession is clearly identified for $q \geq 6$ and $\chi_p \geq 0.2$ or lower mass ratio but higher precession spin. At the same time, when precession is inferred clearly, the amount of precession is consistently underestimated in comparison to the true, injected value.}
\label{tab:chip}
\end{table*}

\begin{figure}[t!]
\includegraphics[width=\columnwidth]{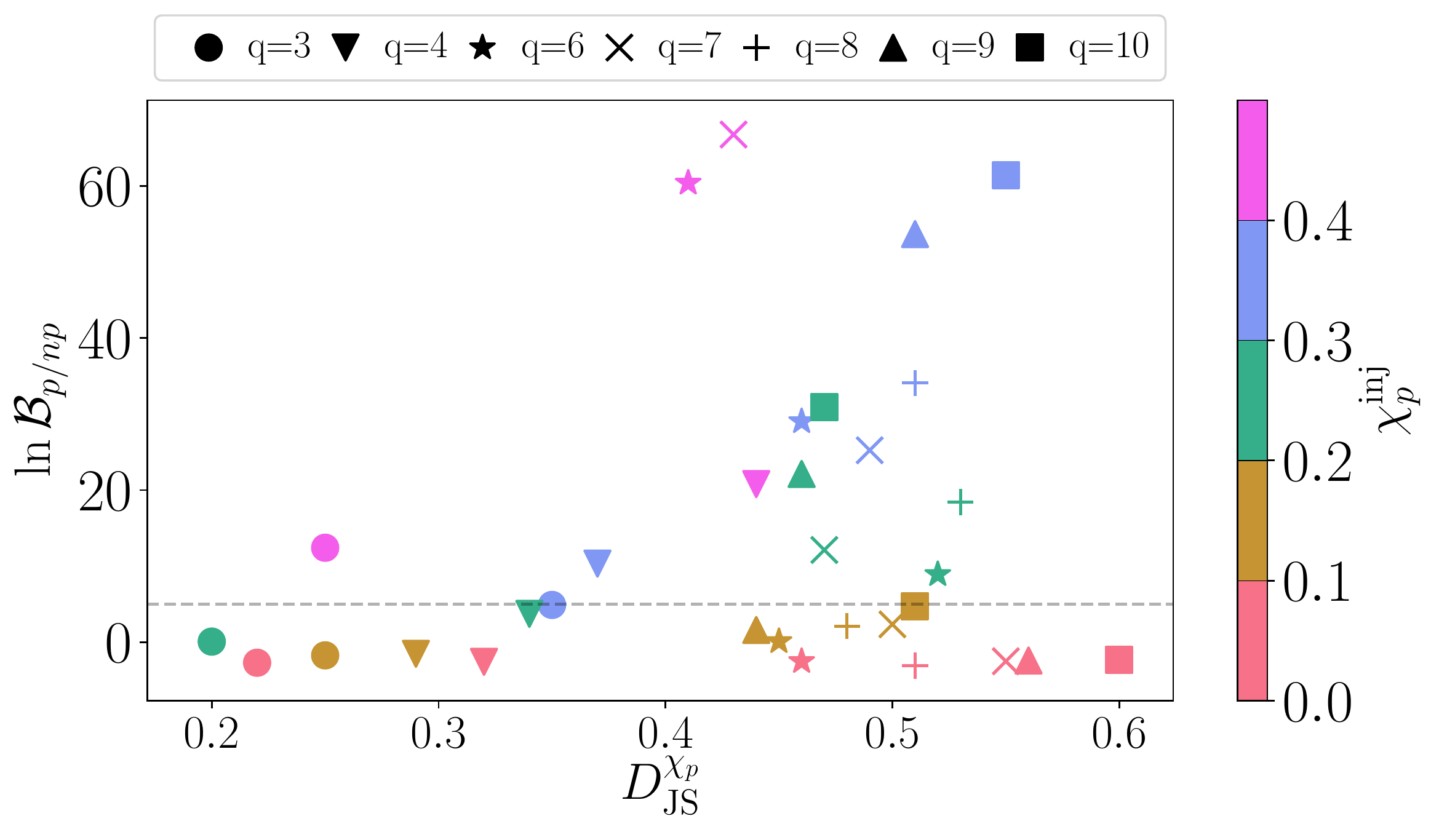}
\caption{Bayes factor vs. the Shannon-Jenson divergence for all binaries considered here. The dashed horizontal line indicates a Bayes factor strongly favouring the precessing signal hypthesis. }
\label{fig:DJS}
\end{figure}

In addition to a Bayes factor in favour of precession, further evidence can be obtained directly from the inferred posterior distribution of $\chi_p$.
We report two information gain measures, $D^{\chi_p}_{\rm JS}$ and $D^{\chi_p}_{\rm KL}$ as defined in Eqs.~\eqref{eq:DKL} and \eqref{eq:DJS}. Due to the correlation between $\chi_p$ and $\chi_{\rm eff}$, we condition the $\chi_p$ prior on the $\chi_{\rm eff}$ posterior via rejection sampling following Ref.~\cite{LIGOScientific:2018mvr}. These measures encapsulate how different the inferred posterior distribution of $\chi_p$ is in comparison to its prior distribution. Figure~\ref{fig:DJS} shows the Shannon-Jensen divergence for $\chi_p$ vs. the Bayes factor; the equivalent representation of $D_{\rm KL}^{\chi_p}$ can be found in Fig.~\ref{fig:DKL} in App.~\ref{sec:appA}. The numerical values are reported in Tab.~\ref{tab:divs} also in App.~\ref{sec:appA}.

Focusing first on the normalised Jensen-Shannon divergence $D^{\chi_p}_{\rm JS} \in [0,1]$ (Fig.~\ref{fig:DJS}), we observe two general trends with minor fluctuations: (i) the divergence increases with mass ratio for all values of $\chi_p$, and (ii) for $q \geq 6$ $D_{\rm JS}$ decreases as $\chi_p$ increases. In all cases we find that information has been gained and for $q\geq 6$ the gain is $> 0.4$. Similar trends are observed for the KL-divergence for all values of $\chi_p$. While these divergence measures are indicative of an appreciable difference between the prior and posterior distribution, on their own they are not enough to state whether or not precession has been identified. From Fig.~\ref{fig:DJS}, however, we notice clearly that non-precessing or mildly precessing signals consistently disfavour the precessing hypothesis and have a large information gain that increases with the mass ratio. 

%In fact, the Bayes factor indicates that despite large values of $D_{\rm JS} (D_{\rm KL})$, the precessing signal hypothesis is not necessarily favoured. The divergences are only meaningful when interpreted together with the moments of the posterior distributions.

% FIGURE DUMP - move as needed

\begin{figure}[t!]
\includegraphics[width=\columnwidth]{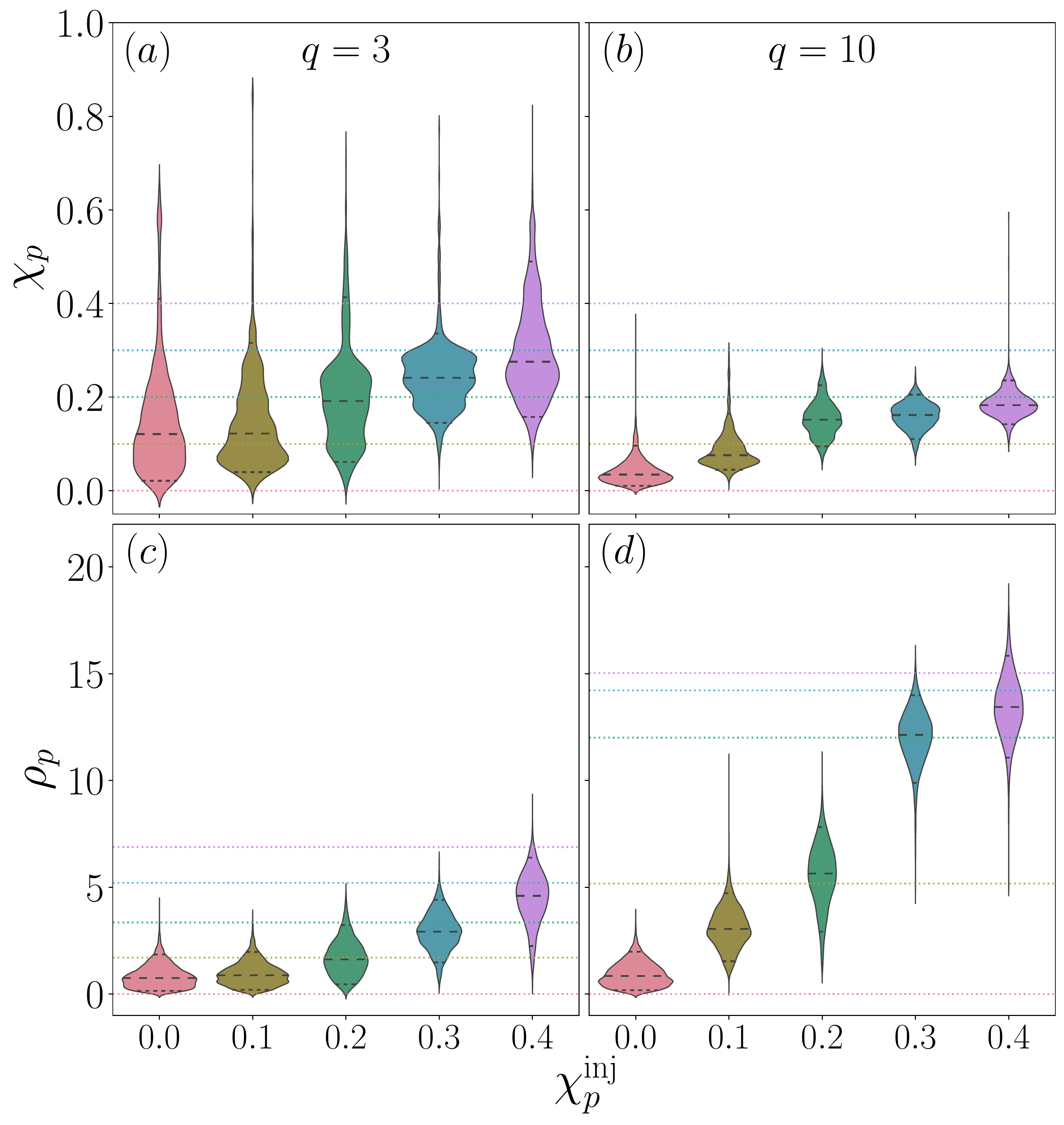}
\caption{Posterior distributions for $\chi_p$, (a) and (b), and $\rho_p$, (c) and (d), for mass ratios $q=3$ and $q=10$ for different amounts of precession (indicated on the $x$-axis). The black dashed and dotted lines within the shaded area indicate the median and $90\%$ CI, respectively. The coloured dotted lines show the value of $\chi_p$ and $\rho_p$ for each injection.}
\label{fig:posteriors}
\end{figure}

This becomes further evident when contrasting the divergences with the median and 90\% CI of the $\chi_p$-posterior distributions given in Tab.~\ref{tab:chip}. For example, we consistently find a $D^{\chi_p}_{\rm KL} \gtrsim 1$ for the nonspinning and therefore nonprecessing binaries and, in combination with the median and 90\% CI, we find that these binaries are correctly identified as nonprecessing or, at worst, as very mildly precessing. Furthermore, we find that the more asymmetric the mass ratio and the larger the intrinsic precession effects, the tighter and more confident the constraints that can be placed on $\chi_p$ are. This is also shown in the top row of Fig.~\ref{fig:posteriors} for binaries with mass ratios $q=3$ and $q=10$, the results for the other mass ratios can be found in Fig.~\ref{fig:posteriors_all} in Appendix~\ref{sec:appA}. In particular, we find that a non-vanishing $\chi_p$ can be constrained away from zero with increasing significance as the mass ratio increases. For a true $\chi_p \geq 0.3$ we find that $\chi_p < 0.08$ is excluded for all mass ratios at 99\% CI; for mass ratios $q > 4$, $\chi_p < 0.07$ is excluded at 99\% CI already for true $\chi_p$-values of $0.2$.
This is not surprising as precession effects become more pronounced in this regime.

We notice, however, that while the divergence from the prior increases and the width of  90\% CI shrinks with increasing $\chi_p$, the recovery of $\chi_p$ is significantly biased (see third column in Tab.~\ref{tab:chip}). For all mass ratios and values of $\chi_p \geq 0.2$, the amount of precession is consistently \emph{underestimated}; only systems with low $\chi_p$ show small positive biases. Furthermore, for all configurations with $q \geq 6$ and $\chi_p \geq 0.2$ the true value of $\chi_p$ lies outside the 90\% credible interval (see Fig.~\ref{fig:posteriors_all} in App.~\ref{sec:appA}), indicating that systematic modelling errors between \texttt{SEOBNRv4PHM} and \texttt{IMRPhenomPv2} dominate over statistical uncertainty.
Similarly, the posterior quantile $Q(\chi_p)$ as given by Eq.~\eqref{eq:quantile} reaffirms the appreciable underestimation of $\chi_p$ for $q \geq 4$ and $\chi_p \geq 0.2$ but additionally tells us about the skew of the inferred $\chi_p$-distribution. 
The observed biases in $\chi_p$ are perhaps not surprising given the differences between the injected waveform model and the one used for parameter recovery. Our results show that systematic modelling errors can affect the accuracy of spin measurements already at current detector sensitivities and relatively moderate SNRs. 
Consistent with the results obtained for GW190814, however, the absence of precession, i.e., $\chi_p \simeq 0$, is unlikely to be misidentified even for moderate inclinations. Our results indicate that precession (or the absence thereof) is robustly identified in such NSBH-like asymmetric binaries at reasonable SNRs and inclinations.

\begin{figure}[t!]
\includegraphics[width=\columnwidth]{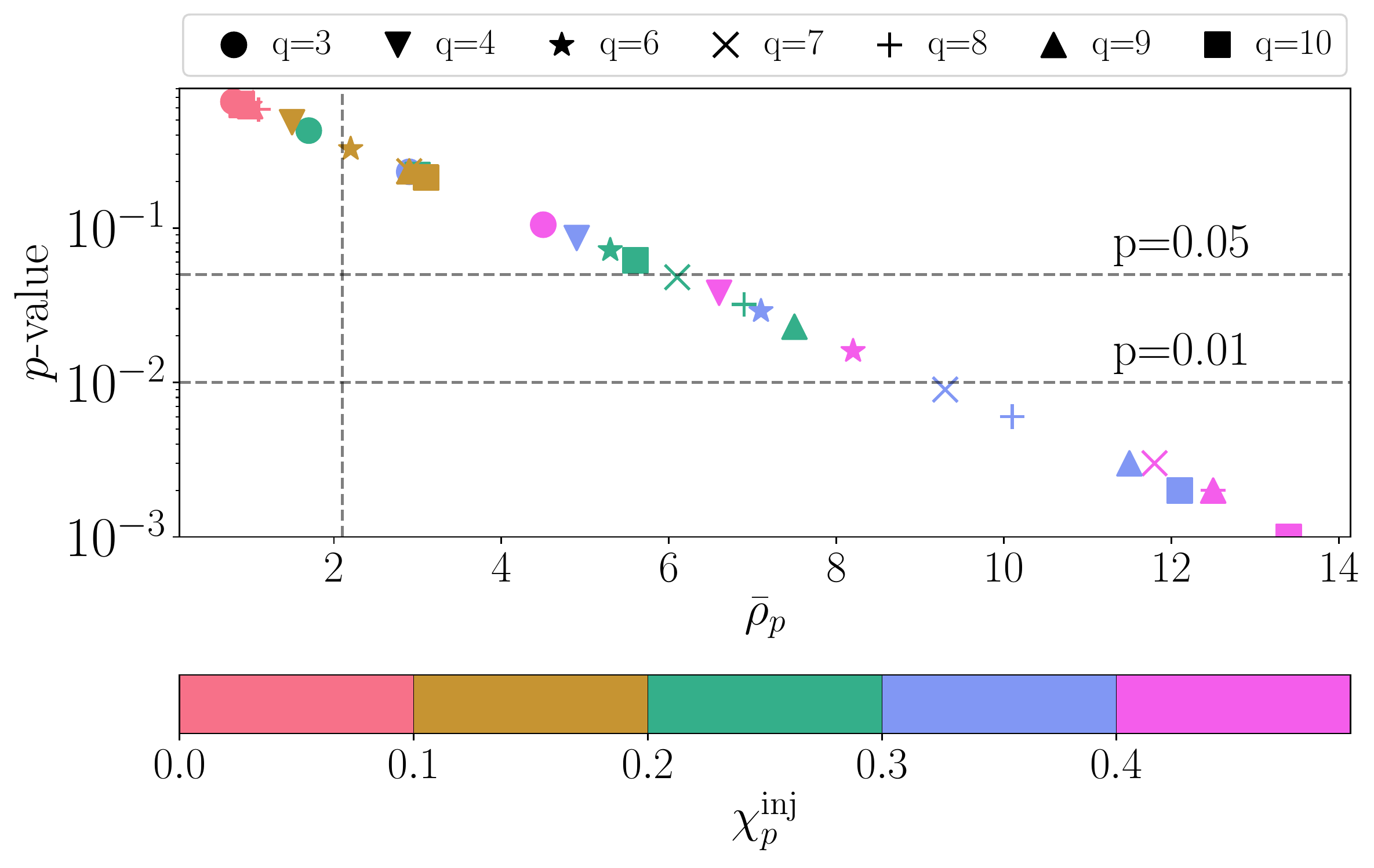}
\caption{One-sided $p$-value of $\rho_p$ as a function of the recovered mean precessing SNR $\bar{\rho}_p$. The dashed vertical line indicates $\rho_p=2.1$ and the dashed horizontal lines a $p$-value of $0.05$ (moderate significance) and $0.01$ (strong significance) respectively.}
\label{fig:pval}
\end{figure}

In addition to the fully Bayesian analysis, we now look at the distributions of the frequentist measure $\rho_p$ for all configurations (see Sec.~\ref{sec:rhop}). For each binary we compute the $\rho_p$-distribution from the posterior samples of the Bayesian analysis using \texttt{PESummary}~\cite{2020arXiv200606639H}. The distributions for $q=3$ and $q=10$ are shown in the bottom panels of Fig.~\ref{fig:posteriors}, the results for the other mass ratios can be found in Fig.~\ref{fig:posteriors_all} in App.~\ref{sec:appA}. The coloured horizontal lines indicate the injected $\rho_p$-value for each value of $\chi_p$. 
We observe trends similar to $\chi_p$: (i) the more asymmetric the mass ratio and the larger $\chi_p$, the more likely that $\rho_p$ exceeds the threshold of $2.1$; (ii) $\rho_p$ is always underestimated except for the nonspinning cases, where it is overestimated. 

To quantify the statistical significance of the inferred $\rho_p$, we compute the $p$-value for its mean relative to a $\chi^2$-distribution with two degrees of freedom, which is the distribution expected in the absence of precession~\cite{Fairhurst:2019har}; the smaller the $p$-value, the more significant the deviation from the non-precessing distribution.  
Figure~\ref{fig:pval} shows the $p$-value as a function of the recovered mean $\bar{\rho}_p$, where the two horizontal lines indicate a $p$-value of $0.05$ (moderate significance) and $0.01$ (strong significance) respectively.
The most significant $p$-values are obtained only for $\chi_p \geq 0.3$ and $q \geq 7$. We find the $\rho_p$-results to be consistent with the results from the fully Bayesian analysis, but they do not provide any additional information or further constraining power. In particular, the $p$-value statistic suggests that the two-harmonics threshold precessing SNR of $2.1$ is too low in the presence of systematic errors. 
The means, $1$-$\sigma$ variances and $p$-values are given in Tab.~\ref{tab:rho_p} in App.~\ref{sec:appA}. 

\begin{figure}
    \centering
    \includegraphics[width=\columnwidth]{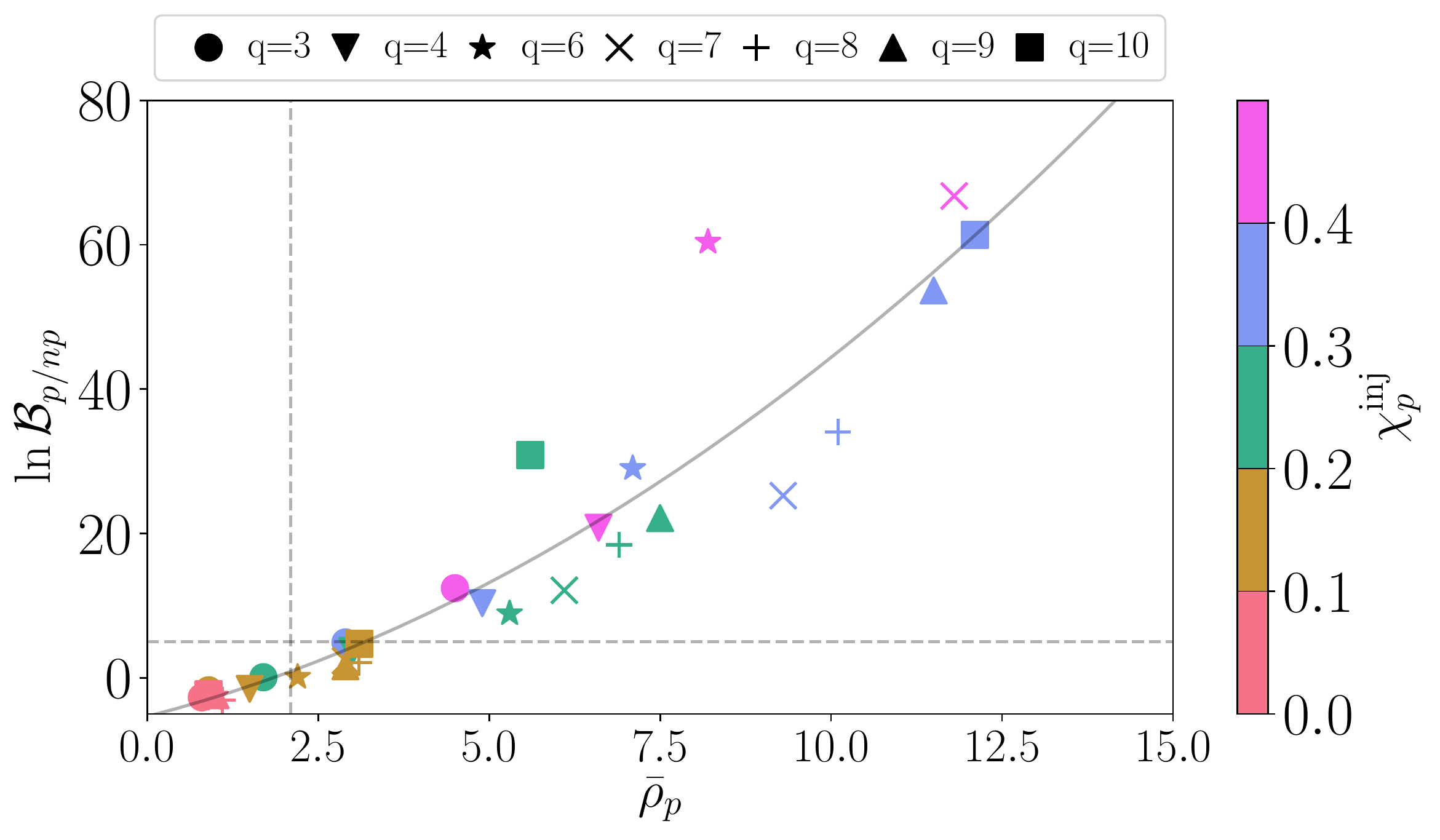}
    \caption{The precessing vs non-precessing Bayes factor for all binaries as a function of the mean recovered precessing SNR $\bar{\rho}_p$. The grey graph indicates the approximate quadratic dependence, $\ln \mathcal{B}_{p/np} \propto \bar{\rho}_p^2$.}
    \label{fig:BFrhop}
\end{figure}

In Fig.~\ref{fig:BFrhop} we also show the precessing vs. non-precessing Bayes factor as a function of the recovered precessing SNR $\bar{\rho}_p$. We find that $\ln \mathcal{B}_{\rm p/np} \propto \bar{\rho}_p^2$. Similar to the $p$-value results, we see that confident statements about the presence of precession are restricted to larger mass ratios and in-plane spin values with a $\bar{\rho}_p$-value of $2.1$ being too low a detection threshold in the presence of waveform systematics. By mapping the recovered precessing SNR to the Bayes factors, it will be possible to estimate the measurability of precession whilst avoiding additional parameter estimation runs. However, this requires a detailed characterisation of the mapping and the impact of waveform systematics. 

%%%%%%%%%%%%%%%%%% MASSES
%%%%%%%%%%%%%%%%
\subsection{Mass--Spin Degeneracy}
\label{sec:mass-improvement}
%%%%%%%%%%%%%%%%
\begin{figure*}[t!]
\includegraphics[width=\columnwidth]{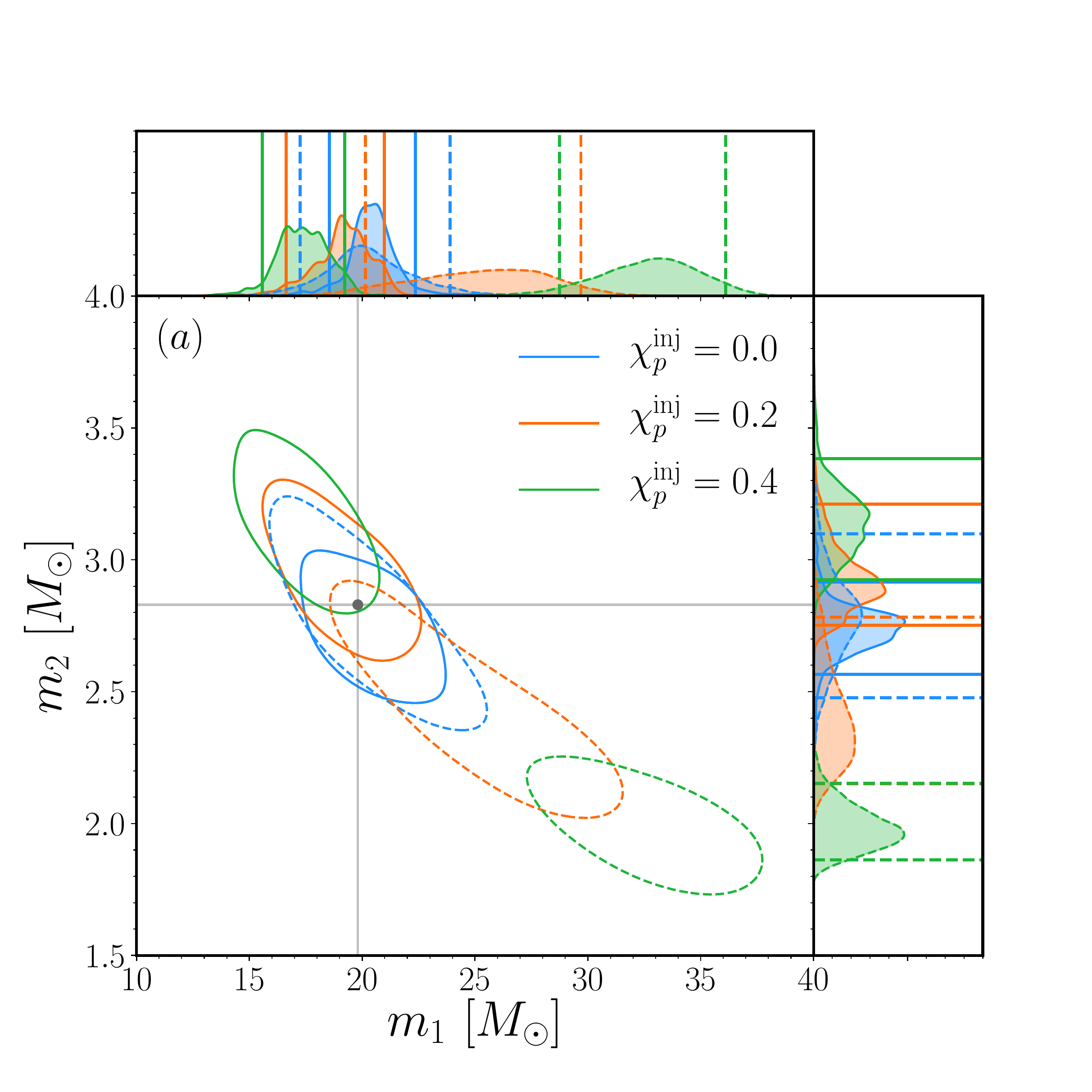}
\includegraphics[width=\columnwidth]{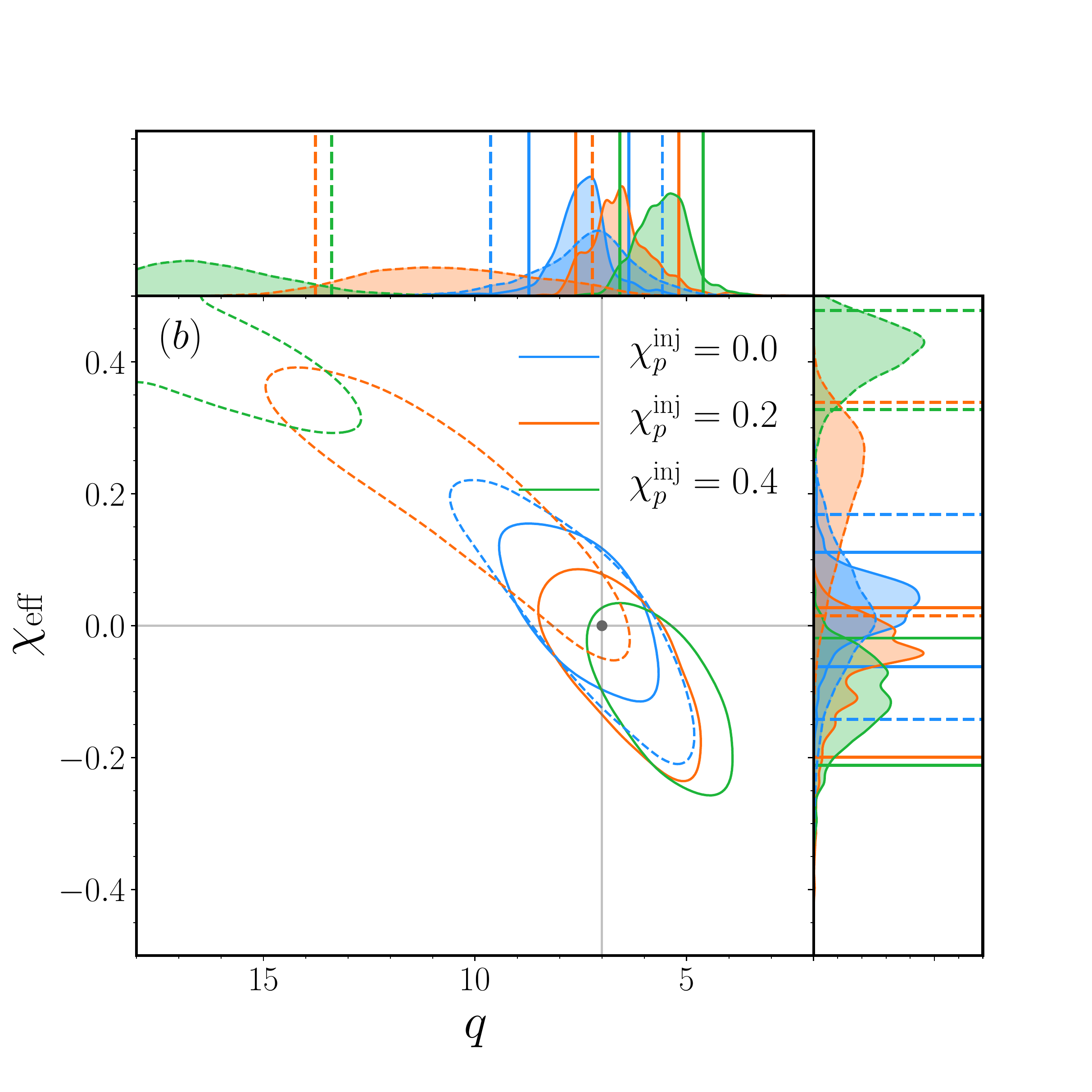}
\caption{One-dimensional and joint posterior distributions for the component masses, panel (a), and the mass ratio and effective aligned spin, panel (b), for a set of $q=7$ binaries as inferred using the \texttt{IMRPhenomPv2} (solid) and \texttt{IMRPhenomD} (dashed) waveform models. As we increase the amount of precession in the injected signal, we find a significant increase in the bias of the inferred masses towards smaller values when recovering with the non-precessing approximant. As can be seen, spin-precession breaks the mass-spin degeneracy in the $q-\chi_{\rm eff}$ plane \cite{Vecchio:2003tn}, allowing a tighter localisation of both the mass and the spin. }
\label{fig:mass-spin-deg-q-chi-eff}
\end{figure*}

\begin{figure*}[t!]
\includegraphics[width=0.67\columnwidth]{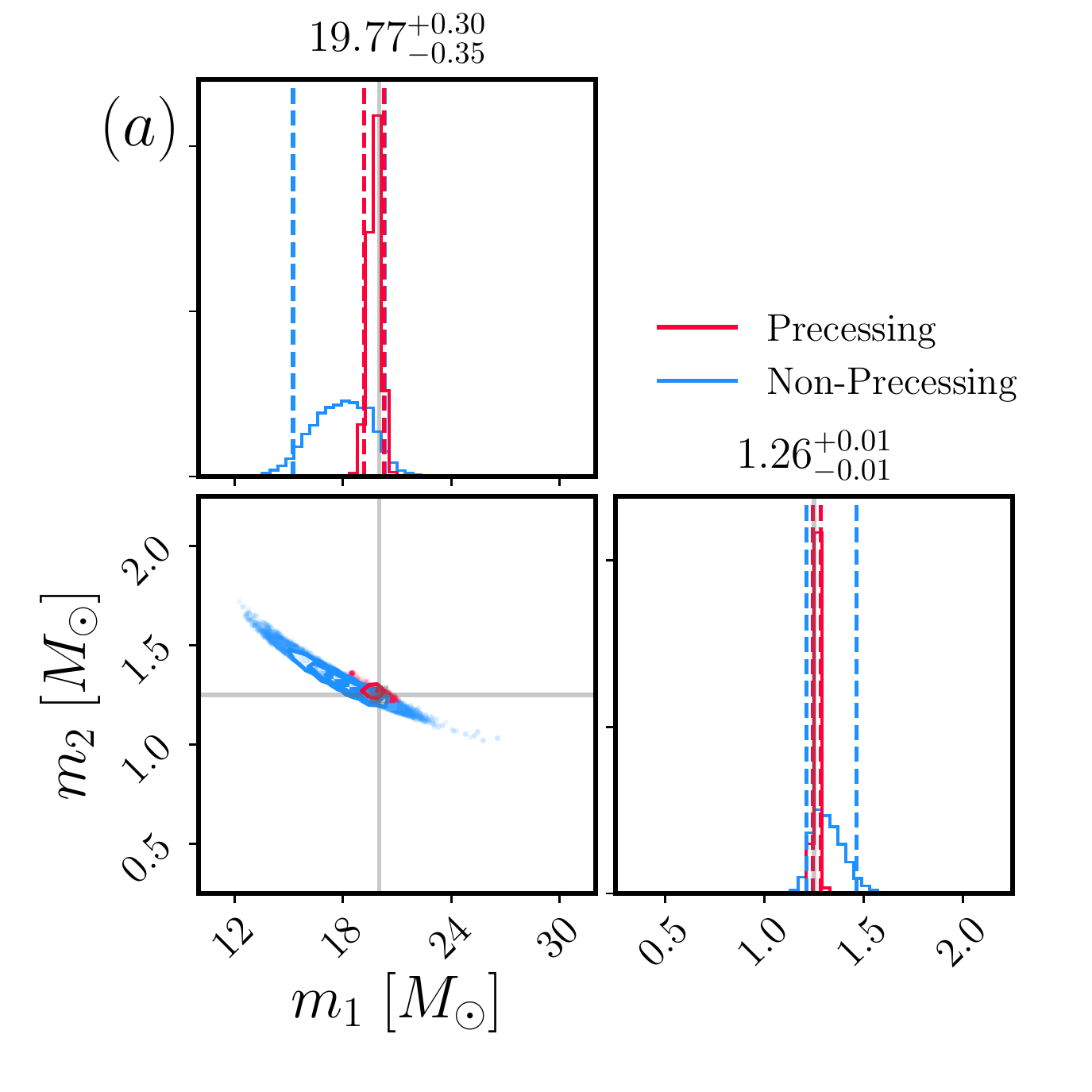}
\includegraphics[width=0.67\columnwidth]{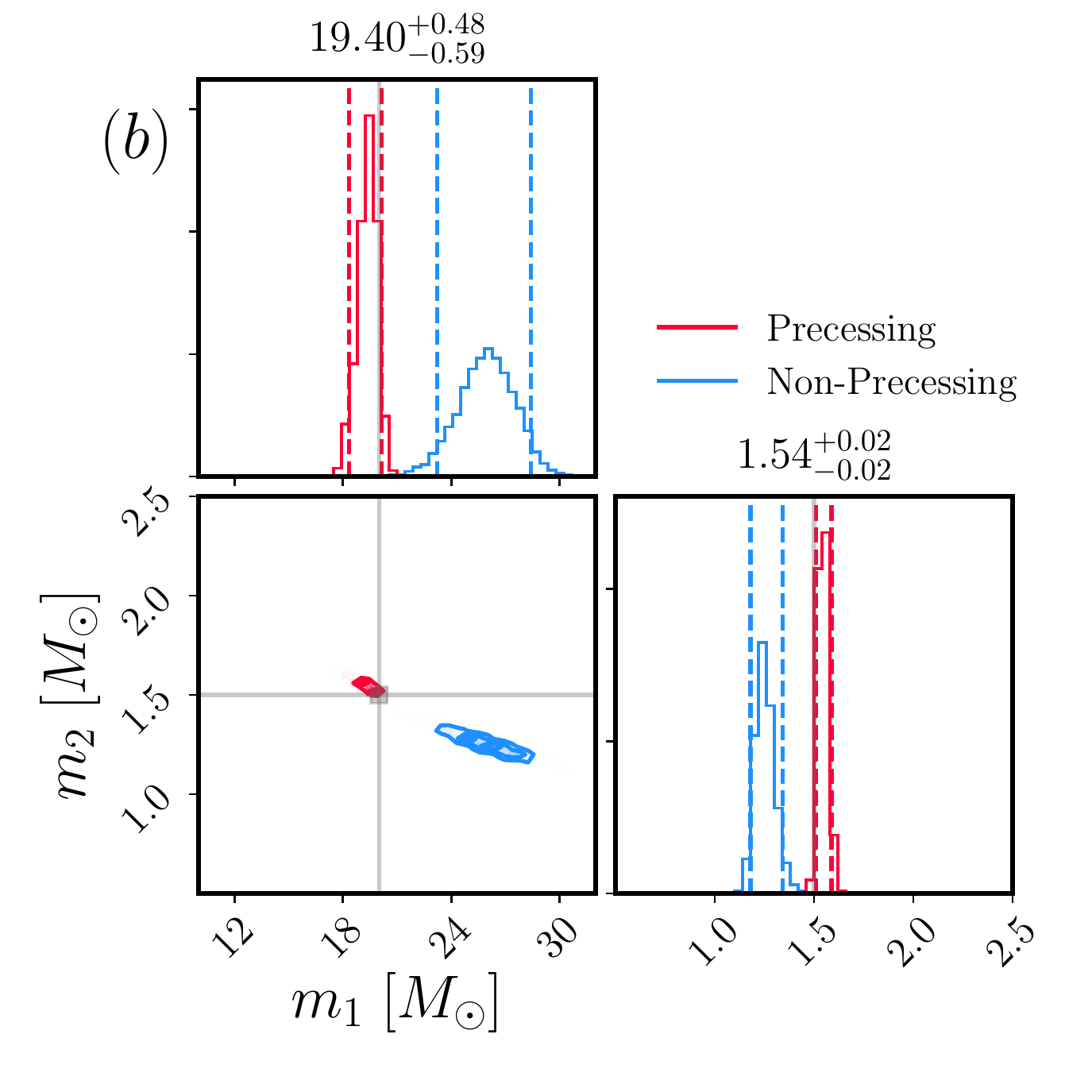}
\includegraphics[width=0.67\columnwidth]{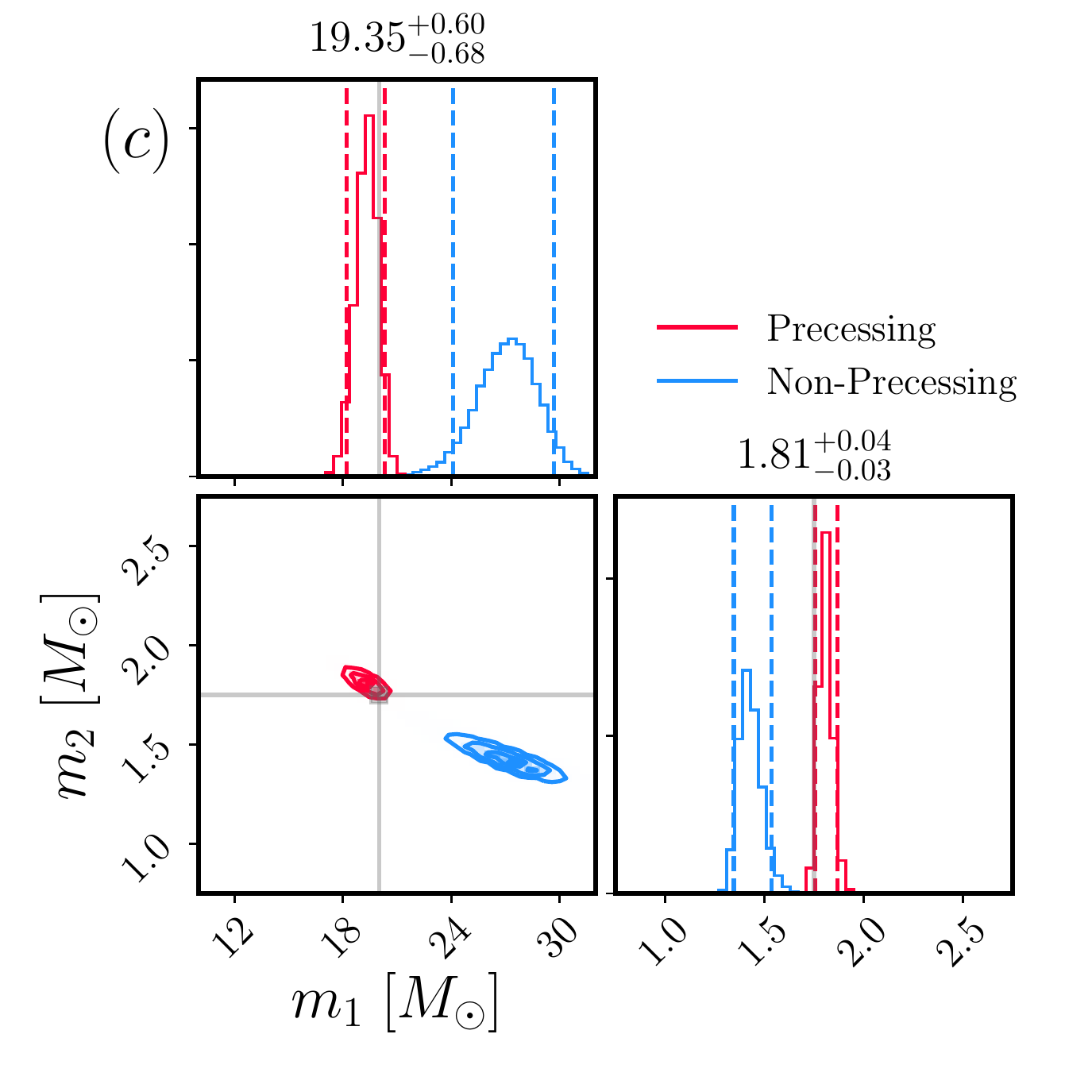}
\includegraphics[width=0.67\columnwidth]{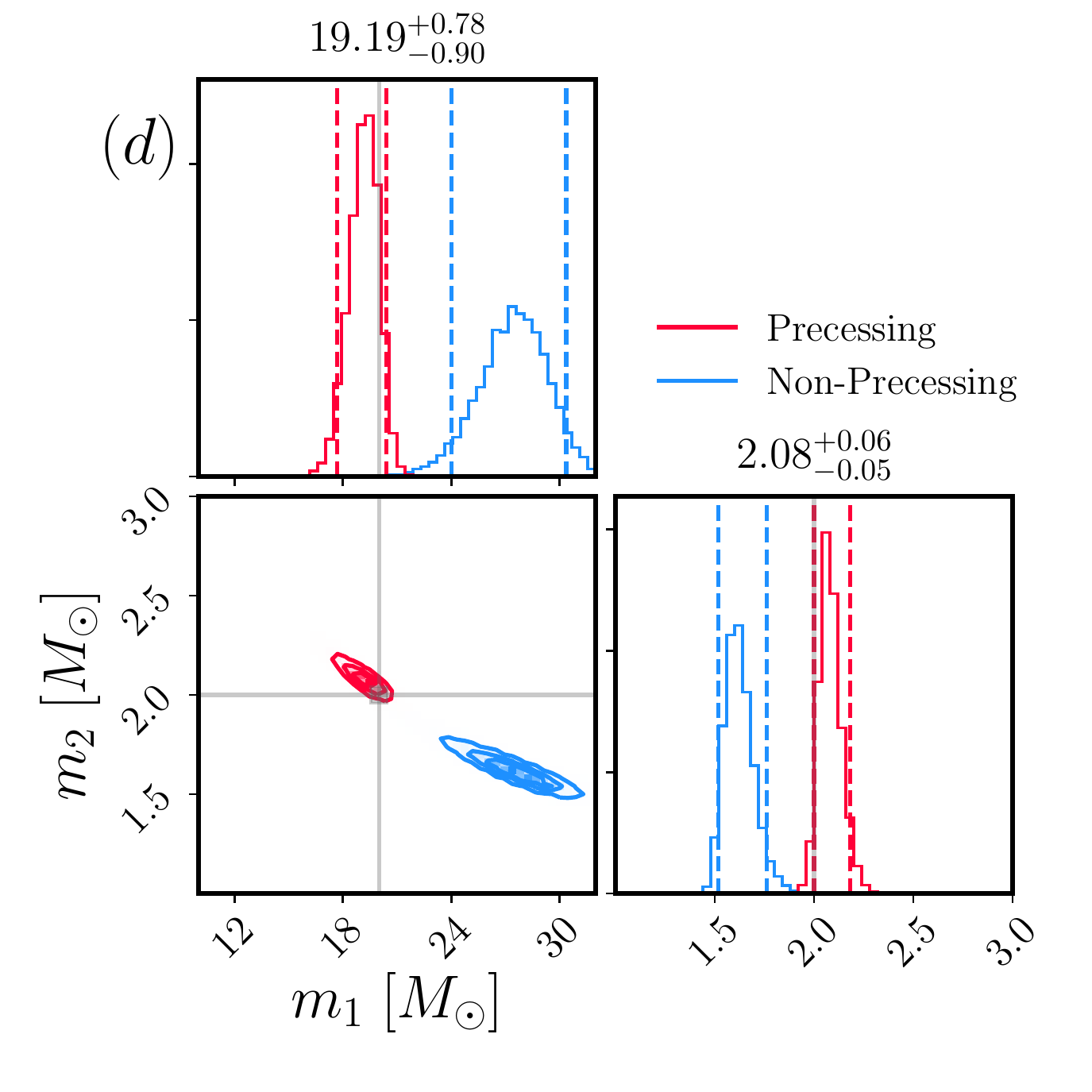}
\includegraphics[width=0.67\columnwidth]{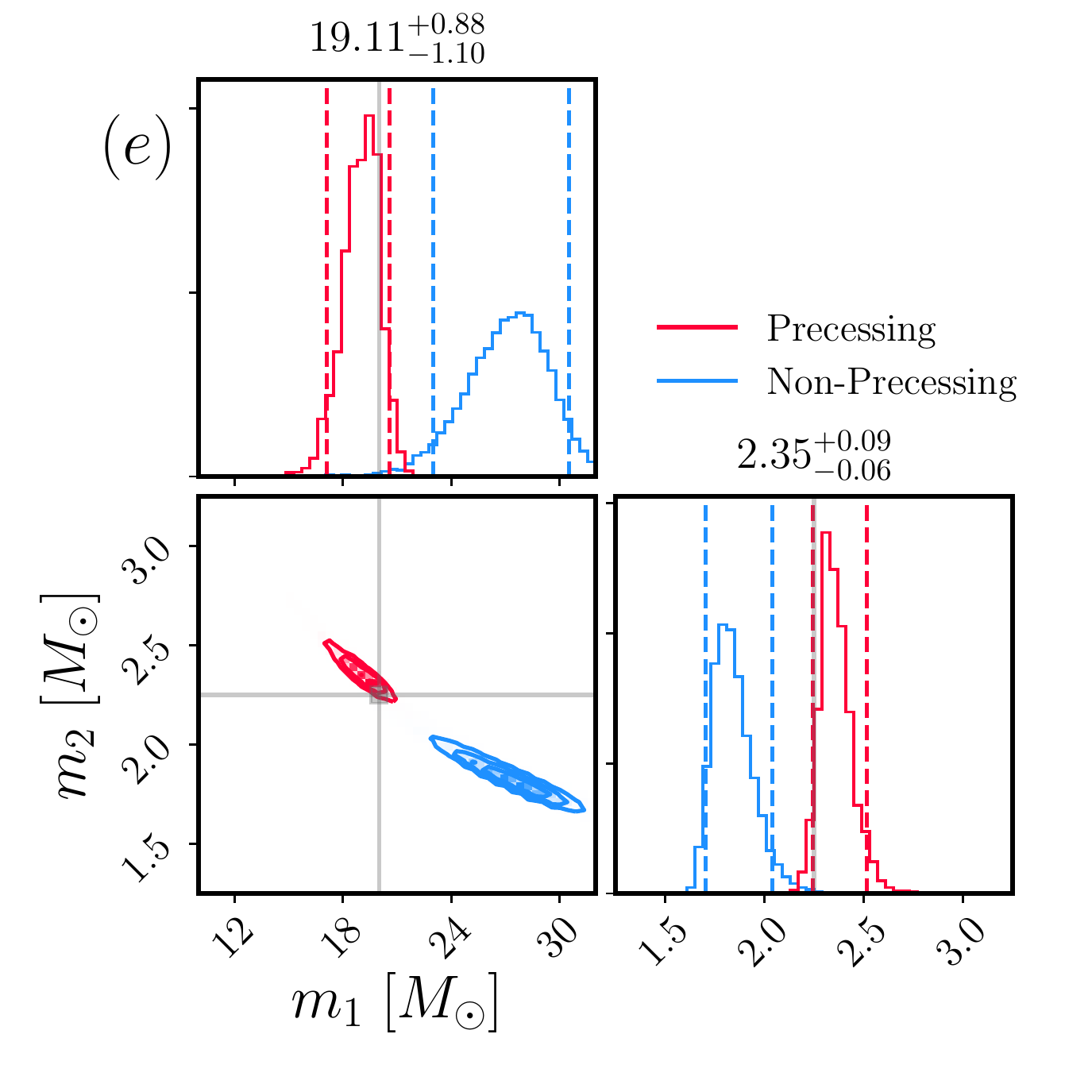}
\includegraphics[width=0.67\columnwidth]{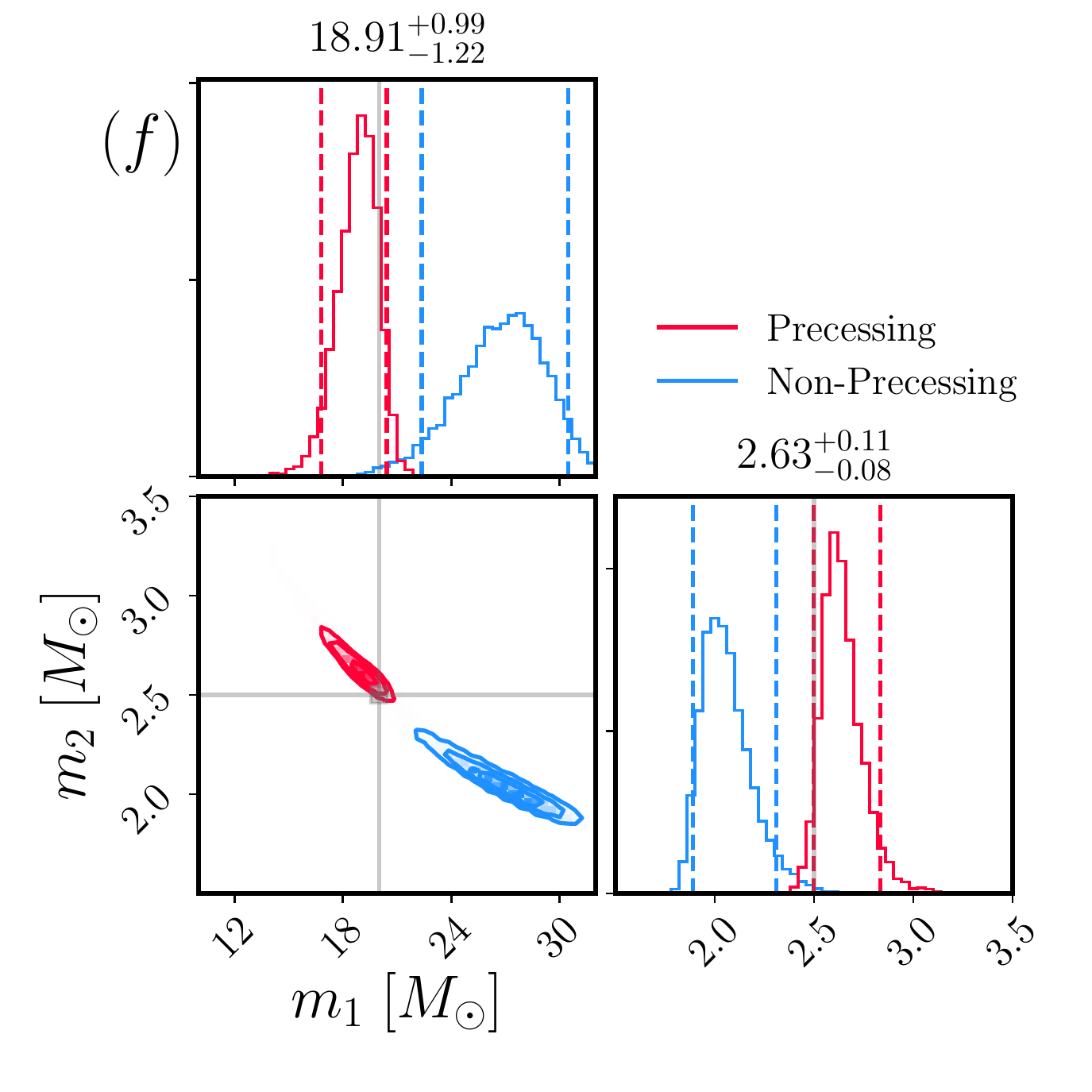}
\includegraphics[width=0.67\columnwidth]{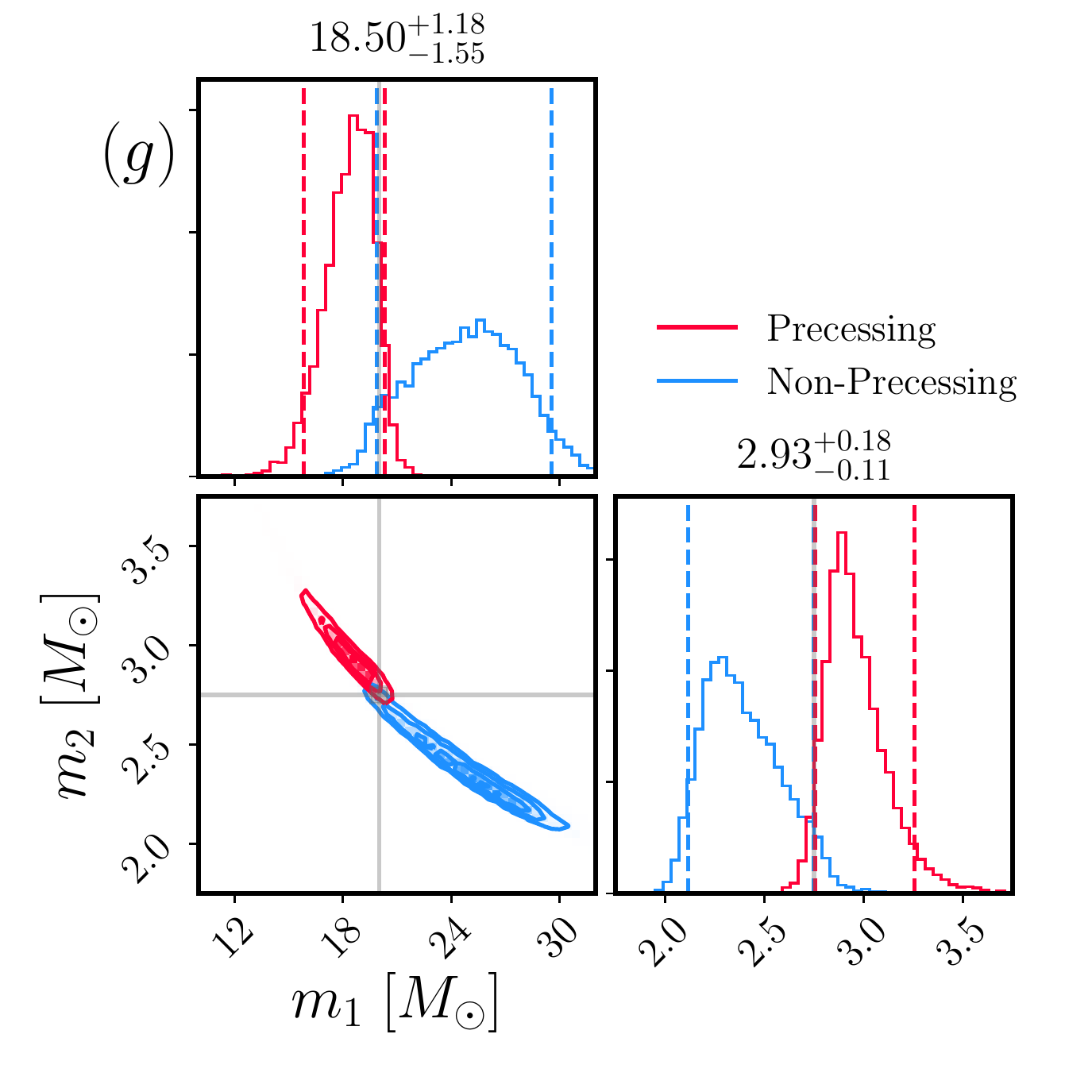}
\includegraphics[width=0.67\columnwidth]{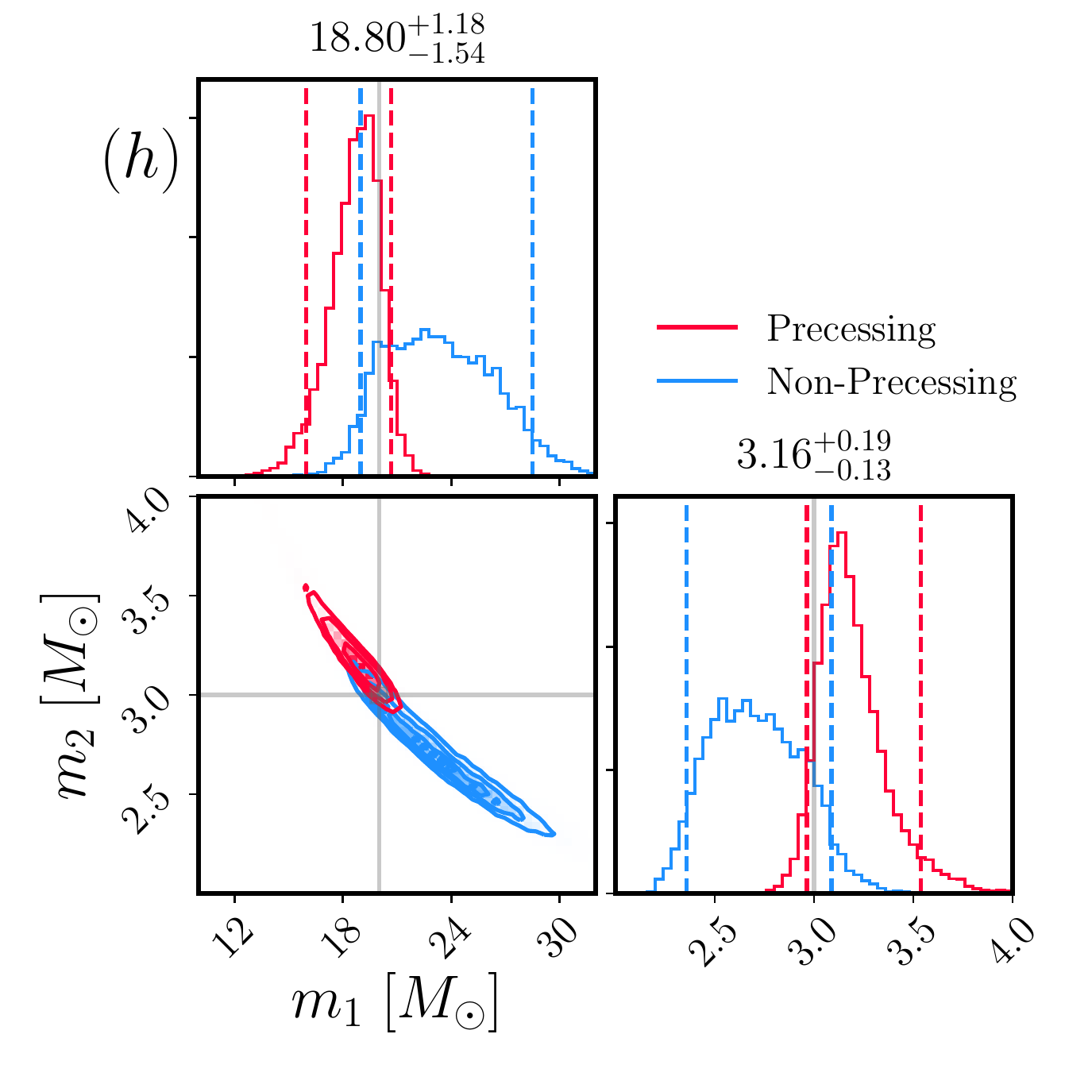}
\caption{One-dimensional and joint mass posteriors for the NSBH-like series of injections. We keep the mass of the primary fixed at $m_1 = 20 M_{\odot}$ and vary the mass of the secondary in the range $m_2 \in \left[ 1.25, 3.00 \right] M_{\odot}$. The SNR of the binary is fixed at $\rho = 30$ with $\chi_{\rm eff} = 0$ and $\chi_p = 0.2$. Spin precession breaks the mass -- spin degeneracy leading to significant improvements in the recovered mass parameters. The non-precessing templates (blue) demonstrate a systematic bias towards heavier primary and lighter secondary masses compared to the precessing templates (red). The grey lines denote the injected component masses. The values reported are the $90\%$ from the precessing model \texttt{IMRPhenomPv2.}}
\label{fig:mass_1_mass_2}
\end{figure*}

Accurate measurement of the component masses is of vital importance in determining the astrophysical nature of low-mass compact objects. This is particularly important for NSBH-like binaries, where there is likely to be no EM counterpart and no discernible information regarding the tidal deformability of the lighter companion \cite{Pannarale:2014rea}. In this section, we assess the confidence to which we can measure the secondary mass in high mass ratio binaries. We focus on two scenarios. In the first scenario, we highlight how the biases in the inferred component masses become progressively worse as we increase the amount of precession in the system. In the second scenario, we consider an astrophysically motivated series in which we fix the mass of the primary and vary the mass of the secondary such that it spans a range of plausible neutron star masses \cite{Abbott:2018wiz, Abbott:2020uma,Ozel:2010apj}. 

In Fig.~\ref{fig:mass-spin-deg-q-chi-eff}, we show the one-dimensional and joint posteriors for the source frame component masses (left panel) for a $q=7$ binary with $\mathcal{M}_c = 6 M_{\odot}$, $\chi_p \in \lbrace 0, 0.2, 0.4 \rbrace$ and all other extrinsic parameters fixed to the values reported in Tab.~\ref{tab:injparams}. We show both the precessing (solid) and non-precessing (dashed) posteriors. By neglecting spin-precession in the recovery waveform model, we find significant biases in the inferred component masses as the magnitude of the in-plane spin is increased. For the most strongly precessing configurations considered here, $\chi_p = 0.4$, the bias in the primary mass is $\Delta m_1 \simeq +13 M_{\odot}$ and in the secondary $\Delta m_2 \simeq -0.88 M_{\odot}$, respectively. In particular, this example demonstrates how a compact object with mass $m_2 = 2.83 M_\odot$, which is significantly heavier than the most massive NS observed to date~\cite{Cromartie:2019kug}, would be misidentified as having a mass of $\simeq 1.95 M_{\odot}$ if spin-precession effects were neglected. Similarly, in the right panel of Fig.~\ref{fig:mass-spin-deg-q-chi-eff}, we highlight how spin-precession breaks the $q-\chi_{\rm eff}$ degeneracy \cite{Vecchio:2003tn,Lang:1900bz,Chatziioannou:2014coa}. 

For the second series, the mass of the primary is fixed to $m_1 = 20 M_{\odot}$ and the secondary mass varies from $m_2 = 1.25 M_{\odot}$ to $m_2 = 3.00 M_{\odot}$. Here, we allow for a small but non-negligible amount of precession with $\chi_p=0.2$. The results are shown in Fig.~\ref{fig:mass_1_mass_2}. For all binaries considered in this series, the posteriors obtained using \texttt{IMRPhenomPv2} are demonstrably less biased, with the true injected masses being always contained within the 90\% CI. In addition, the posteriors are tighter than the posteriors inferred using \texttt{IMRPhenomD}. As we increase $m_2$, we increase $\mathcal{M}_c$ but decrease the mass ratio. Consequentially, we find that the \texttt{IMRPhenomD} posteriors become progressively less biased but the posteriors widths become broader. We observe that the non-precessing approximant significantly underestimates the mass of the secondary for nearly all binaries considered, leading to stronger support for masses that are consistent with known theoretical bounds on the maximum NS mass. In contrast, as \texttt{IMRPhenomPv2} is recovering almost unbiased mass estimates, with the true injected value always lying towards the lower 90\% CI, the posterior support for plausible neutron star masses is significantly reduced. Of particular note is the $m_2 = 2.75 M_{\odot}$ injection, falling just above current causal bounds on the NS mass, where \texttt{IMRPhenomPv2} demonstrates little posterior support for $m_2 < 2.75 M_{\odot}$ whereas \texttt{IMRPhenomD} has posterior support down to $2 M_{\odot}$; also see Fig.~\ref{fig:mass_2_oned_pos} for a comparison of the inferred one-dimensional posterior distributions for $m_2$. Whilst only a preliminary study on a single set of injections, these results serve to highlight the importance of including spin-precession in our waveform models when making inferences about the nature of the secondary compact object \cite{Chatterjee:2016thb,Chen:2019aiw}. We note that misidentifying a light BH as a heavy NS will introduce significantly less bias in inferred NS parameters than misidentifying a BH as a light NS \cite{Chen:2019aiw}. In such scenarios, the use of non-precessing approximants for parameter estimation could introduce non-trivial biases in the inferred population properties, including inferences on the NS equation of state \cite{Essick:2019ldf,LIGOScientific:2019eut,Wysocki:2020myz}.

As a caveat to the analysis discussed here, we neglect the role that tidal effects and tidal disruption could have on the morphology of a NSBH waveform \cite{Shibata:2009cn,Shibata:2011jka,Foucart:2010eq,Kyutoku:2010zd,Kyutoku:2011vz,Foucart:2012vn,Foucart:2012nc,Foucart:2019bxj}. As we move to larger mass-ratios, the occurrence of tidal disruption becomes increasingly unlikely and the waveform begins to closely resemble that of a BBH with the high-frequency behaviour of the amplitude being governed by the ringdown of the primary BH \cite{Foucart:2013psa}. For more comparable mass ratios, significant tidal disruption of the NS can take place and the amplitude becomes exponentially suppressed at high frequencies. Several non-precessing waveform models have incorporated such effects \cite{Lackey:2013axa,Pannarale:2015jka, Thompson:2020nei,Matas:2020wab} but no precessing NSBH waveform models are yet available. The impact of tidal disruption on statistical and systematic uncertainties in non-spinning NSBH binaries has recently been investigated in \cite{Huang:2020pba}, where it was shown that neglecting tidal contributions introduces systematic biases for comparable mass ratios but at these highly asymmetric mass ratios spin effects are expected to be the more important one.

%%%%%%%%%%%%%%% DISCUSSION & CONCLUSIONS
\section{Discussion}
\label{sec:discussion}
%%%%%%%%%%%%%%%
Accurate measurements of the component masses and misaligned spins are of prime importance in understanding the origin and evolution of astrophysical compact binaries. It is therefore imperative that we understand how robust such measurements from GW observations are. In this work, we have re-assessed our ability to discern spin-precession in high mass ratio binaries similar to GW190814 in the current detector era. We have quantified this using Bayesian model selection supplemented by additional Bayesian and frequentist measures. 

Due to the large number of parameters that characterise a precessing compact binary, many studies on the measurability of precession have commonly focused on statements made at the population level \cite{Vitale:2014mka,Littenberg:2015tpa,Talbot:2017yur}. Detailed systematic studies are rare~\cite{Vitale:2016avz, Trifiro:2015zda, Cho:2012ed, Afle:2018slw}. 
Here, we consider a restricted series of injections designed to understand how systematically increasing the amount of precession impacts our ability to make statements on the measurability of precession in GW190814-like, and how the neglect of precession in waveform models leads to non-trivial biases in the inferred component masses, which can have crucial implication for NSBH-like systems. 

Our results show that even small amounts of precession are robustly identified for moderately asymmetric mass ratios $q > 5$. For less unequal masses, larger amounts of precession are required to make robust statements. For all mass ratios we find that model selection alone does not allow to differentiate between a non-precessing binary and a binary with $\chi_p < 0.1$; for small asymmetric mass ratios an even larger amount of precession is required for model selection to discriminate. For all mass ratios $q > 3$ precession with $\chi_p > 0.2$ is robustly measured but biased towards lower values, showing that systematic errors can already be of concern at current detector sensitivities.
As illustrated for the $q=9$ case, we expect that lower SNR signals will need to be more strongly precessing to obtain a Bayes factor high enough to distinguish between the precessing and non-precessing hypothesis. For binaries with higher chirp masses, where fewer precession cycles are detectable, preliminary studies show similar trends but we leave a comprehensive analysis to future work. As for smaller (larger) inclinations, previous work suggests that it will be more difficult (less difficult) to identify precession conclusively~\cite{Abbott:2016wiq}. 

Furthermore, our analysis highlights how even relatively mild amounts of precession can lead to significant biases in the inferred component masses. Systematically increasing the amount of precession in the system leads to a significant over (under) estimation of the primary (secondary) mass when using an aligned-spin approximant. Precession also breaks the mass -- spin degeneracy, and we consequently find that the posterior widths for the component masses inferred using a non-precessing approximant are a factor $\sim 2$ broader than the equivalent posteriors inferred using the precessing approximant. 

In our analyses, we used a fixed inclination and polarization, and systematically varied the mass ratio $q$ and spin precession $\chi_p$. We restricted our analysis to NSBH-like binaries whose chirp mass is consistent with the values reported for GW190814 \cite{GW190814}. Larger studies exploring the full dependence on the sky location, orientation, masses and full spin degrees of freedom will be important but are beyond the scope of this paper. Further, we only consider binaries with $\chi_{\rm eff} \sim 0$, which is consistent with current observations \cite{LIGOScientific:2018mvr} and theoretical modelling of NSBH systems which predicts large spin misalignment for a high fraction of binaries~\cite{Kalogera:1999tq}.

The analyses presented in this paper could be improved by incorporating higher modes (HM)~\cite{OShaughnessy:2014shr,Cotesta:2018fcv,Nagar:2019wds,Garcia-Quiros:2020qpx,Nagar:2020pcj} and improved modelling of precession \cite{Pratten:2020fqn,Pratten:2020ceb,Ossokine:2020v4phm} into the recovery waveform, where we anticipate tighter constraints on the component masses, spins and the orientation of the binary. 
Further, for lower mass ratio binaries ($q\leq 4)$, tidal effects which are not included in our analysis may become important. Since tidal parameters are also correlated with the mass, waveforms that include finite-size effects, tidal disruption and precession will be relevant \cite{Chen:2019aiw}. 

The detection of GW190412 and GW190814 provided the first GW observations of highly asymmetric compact binaries. This has opened a new window onto novel relativistic effects, including spin precession and higher-order modes. As gravitational-wave detectors approach design sensitivity, it will be increasingly important to understand systematic errors in the waveform models and the impact on parameter estimation.

%%%%%%%%%%%%%%%%%%%%%%%%%%%%
\section*{Acknowledgments}
%%%%%%%%%%%%%%%%%%%%%%%%%%%%
We thank Alberto Vecchio and Serguei Ossokine for useful discussions and Richard O'Shaughnessy for comments on the manuscript, and Stephen Fairhurst, Rhys Green, Mark Hannam and Charlie Hoy for providing early access to the code used to calculate $\rho_p$
We are grateful for computational resources provided by Cardiff University, and
funded by STFC grants ST/I006285/1 and ST/V001167/1 supporting the UK Involvement in the Operation of Advanced LIGO.
PS acknowledges NWO Veni Grant No. 680-47-460. 
RB is supported by the School of Physics and Astronomy at the University of Birmingham and the Birmingham Institute for Gravitational Wave Astronomy.
LMT is supported by STFC, the School of Physics and Astronomy at the University of Birmingham and the Birmingham Institute for Gravitational Wave Astronomy. 
This manuscript has the LIGO document number P2000224.

%%%%%%%%%%%%%%% APPENDICES
\appendix

\section{Supplementary Information}
\label{sec:appA}
In addition to the figures and tables in the main text, we provide further details and complementary figures here. 

\begin{figure}[h!]
\includegraphics[width=\columnwidth]{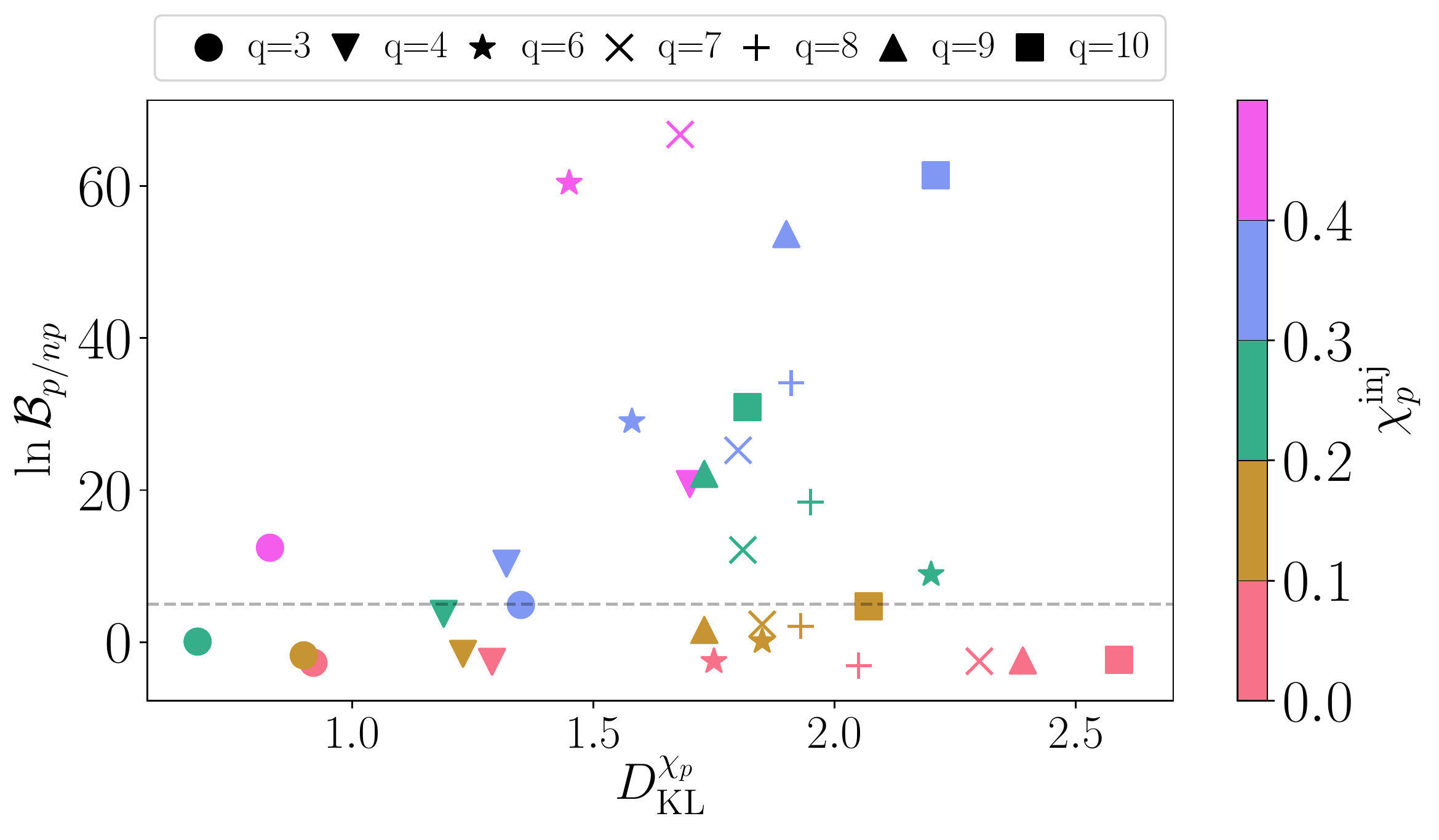}
\caption{Bayes factor vs. the KL-divergence for all binaries considered here. The dashed horizontal line indicates a Bayes factor of $5$, strongly favouring the precessing signal hypothesis. }
\label{fig:DKL}
\end{figure}

Table~\ref{tab:divs} gives the numerical values for the JS- and the KL-divergences for $\chi_p$. Figure~\ref{fig:DKL} is the equivalent of Fig.~\ref{fig:DJS} for the KL-divergence.  

\begin{turnpage}
\begin{figure*}[t!]
\includegraphics[scale=0.30]{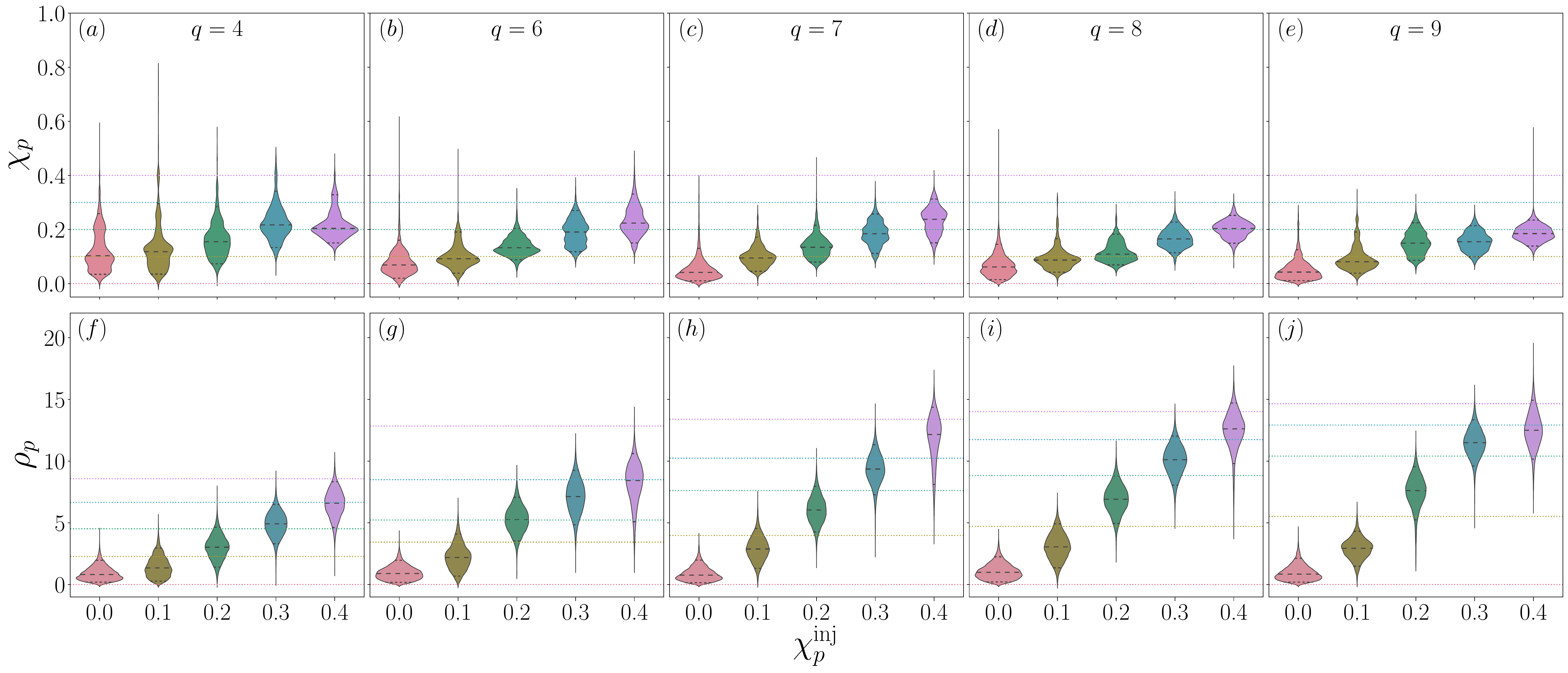}
\caption{One-dimensional posterior distributions for $\chi_p$, (a) to (e), and $\rho_p$, (f) to (j), for mass ratios $q=\{4,6,7,8,9\}$ as a function of increasing $\chi_p$ indicated on the $x$-axis. The lines within the shaded area indicate the median (dashed) and $90\%$ CI (dotted). The coloured dotted lines show the value of $\chi_p$  and $\rho_p$ for each injection.
$\chi_p$ is constrained away from zero with increasing significance as the mass ratio and injected $\chi_p$ increase. 
For all configurations with $q \geq 6$ and $\chi_p \geq 0.2$ the true value of $\chi_p$ lies outside the 90\% credible interval, showing that systematic errors start to exceed the statistical uncertainty. The $\rho_p$-distributions show the same trends, with $\rho_p$ only exceeding the threshold of $2.1$ at the $1$-$\sigma$ level for injections with $q \geq 6$ and $\chi_p \geq 0.2$.}
\label{fig:posteriors_all}
\end{figure*}
\end{turnpage}

Figure~\ref{fig:posteriors_all} is the complement to Fig.~\ref{fig:posteriors} in Sec.~\ref{sec:prec} showing the results for the remaining mass ratios as detailed in Sec.~\ref{sec:inj}. 

In Tab.~\ref{tab:rho_p} we give the numerical values obtained for the mean precessing SNR $\bar{\rho}_p$ and the corresponding $p$-values.

\begin{figure*}[]
\includegraphics[width=0.9\textwidth]{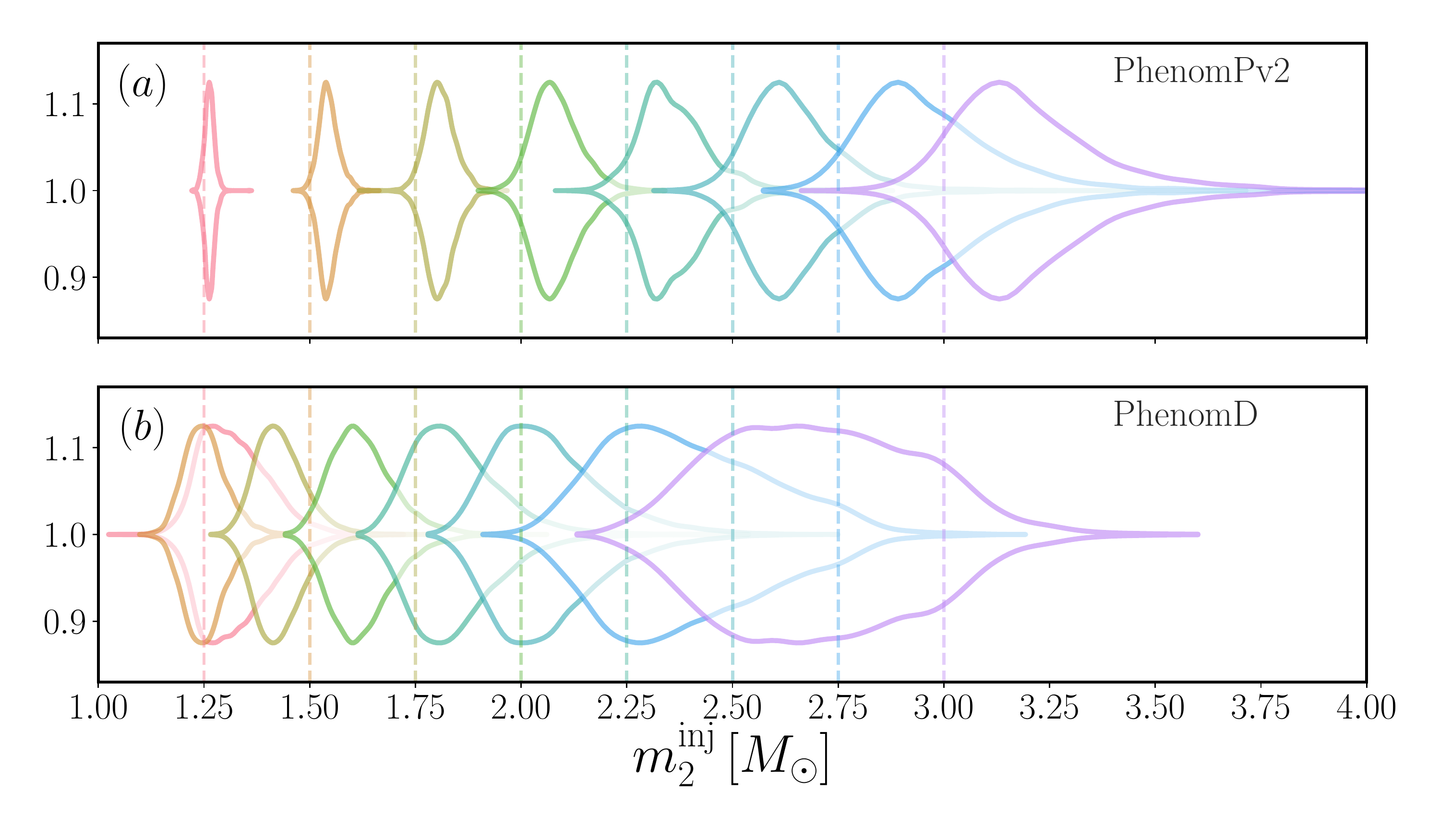}
\caption{One-dimensional posterior distributions for secondary mass as a function of the injected secondary mass (vertical dashed lines). Panel (a) shows the posteriors obtained using \texttt{IMRPhenomPv2} and panel (b) using \texttt{IMRPhenomD} (b). As highlighted in Sec.~\ref{sec:mass-improvement}, spin-precession breaks the mass -- spin degeneracy allowing for tighter constraints on the component masses. For all injections considered here, \texttt{IMRPhenomD} both systematically underestimates the mass of the secondary and has significantly broader posteriors.}
\label{fig:mass_2_oned_pos}
\end{figure*}

In Fig.~\ref{fig:mass_2_oned_pos} we show the one-dimensional posterior distributions of the secondary mass as a function of the injected value. 

% Table with divergences; moved from main
\begin{table}[t!]
\setlength{\tabcolsep}{2.3pt}
\centering
{\renewcommand{\arraystretch}{1.6}
\begin{tabular}{c|a b a b a|b a b a b}
\hline
\hline
& \multicolumn{5}{ c }{$D_{\rm JS}^{\chi_p} (\chi_{\rm eff})$ [bits]} & \multicolumn{5}{ |c }{$D_{\rm KL}^{\chi_p} (\chi_{\rm eff})$ [bits]} \\[0.15cm]\hline
    $\chi_p^{\rm inj}$ & $0.0$ & $0.1$ & $0.2$ & $0.3$ & $0.4$ & $0.0$ & $0.1$ & $0.2$ & $0.3$ & $0.4$ \\[0.05cm]\hline
\hline
     $q=3$ & 0.22 & 0.25 & 0.20 & 0.35 & 0.25 & 0.92 & 0.90 & 0.68 & 1.35 & 0.83 \\[0.07cm]
\hline
     $q=4$ & 0.32 & 0.29 & 0.34 & 0.37 & 0.44 & 1.29 & 1.23 & 1.19 & 1.32 & 1.70 \\[0.07cm]
\hline
     $q=6$ & 0.46 & 0.45 & 0.52 & 0.46 & 0.41 & 1.75 & 1.85 & 2.20 & 1.58 & 1.45 \\[0.07cm]
\hline 
     $q=7$ & 0.55 & 0.50 & 0.47 & 0.49 & 0.43 & 2.30 & 1.85 & 1.81 & 1.80 & 1.68 \\[0.07cm]
\hline 
     $q=8$ & 0.51 & 0.48 & 0.53 & 0.51 & 0.55 & 2.05 & 1.93 & 1.95 & 1.91 & 2.23 \\[0.07cm]
\hline 
     $q=9$ & 0.56 & 0.44 & 0.46 & 0.51 & 0.56 & 2.39 & 1.73 & 1.73 & 1.90 & 2.29 \\[0.07cm]
\hline
     $q=10$ & 0.60 & 0.51 & 0.47 & 0.55 & 0.56 & 2.59 & 2.07 & 1.82 & 2.21 & 2.23 \\[0.07cm]
\hline
\hline
\end{tabular}
}
\caption{Information gain (in bits) between the prior and posterior for $\chi_p$. We show the $D_{\rm JS}$ and $D_{\rm KL}$ divergences at all mass ratios and spins considered. We condition the prior on $\chi_p$ by the posteriors on $\chi_{\rm eff}$.}
    \label{tab:divs}
\end{table}

\begin{table*}[t!]
\setlength{\tabcolsep}{3.pt}
\centering
{\renewcommand{\arraystretch}{1.6}
    \begin{tabular}{c|a b a b a | b a b a b}
    \hline
    \hline 
    & \multicolumn{5}{c}{$\bar{\rho}_p$} & \multicolumn{5}{ |c }{$p$-value} \\[0.15cm]\hline
    $\chi_p^{\rm inj}$ & $0.0$ & $0.1$ & $0.2$ & $0.3$ & $0.4$ & $0.0$ & $0.1$ & $0.2$ & $0.3$ & $0.4$  \\ [0.05cm]\hline
    \hline
    $q=3$ & $0.8 \pm 0.3$ & $0.9 \pm 0.3$ & $1.7 \pm 0.7$ & $2.9 \pm 0.8$ &
$4.5 \pm 1.5$ & 0.657 & 0.622 & 0.427 & 0.231 & 0.105 \\[0.07cm]\hline
    $q=4$ & $0.9 \pm 0.3$ & $1.5 \pm 0.7$ & $3.0 \pm 1.0$ & $4.9 \pm 1.0$ &
$6.6 \pm 1.3$ & 0.634 & 0.483 & 0.220 & 0.086 & 0.038 \\[0.07cm]\hline
    $q=6$ & $1.0 \pm 0.3$ & $2.2 \pm 1.1$ & $5.3 \pm 1.2$ & $7.1 \pm 1.9$ & 
$8.2 \pm 2.8$ & 0.620 & 0.325 & 0.072 & 0.029 & 0.016 \\[0.07cm]\hline
    $q=7$ & $0.9 \pm 0.3$ & $2.9 \pm 1.0$ & $6.1 \pm 1.3$ & $9.3 \pm 1.5$ & 
$11.8 \pm 3.5$ & 0.644 & 0.233 & 0.048 & 0.009 & 0.003 \\[0.07cm]\hline
    $q=8$ & $1.1 \pm 0.4$ & $3.1 \pm 1.1$ & $6.9 \pm 1.4$ & $10.1 \pm 1.5$ &
$12.5 \pm 2.3$ & 0.585 & 0.214 & 0.032 & 0.006 & 0.002 \\[0.07cm]\hline
    $q=9$ & $1.0 \pm 0.4$ & $2.9 \pm 0.7$ & $7.5\pm 1.8$ & $11.5 \pm 1.3$ & 
$12.5 \pm 2.1$ & 0.616 & 0.233 & 0.023 & 0.003 & 0.002 \\[0.07cm]\hline
    $q=10$ & $0.9\pm 0.3$ & $3.1 \pm 0.9$ & $5.6 \pm 2.2$ & $12.1 \pm 1.6$ & 
$13.4 \pm 2.2$ & 0.627 & 0.212 & 0.062 & 0.002 & 0.001 \\[0.07cm]\hline
    \hline 
 \end{tabular}
 }
    \caption{Mean and $1\sigma$-variance of $\rho_p$ and its $p$-value for all binary configurations. The $p$-value is calculated for the mean w.r.t. a $\chi^2$-distribution with two degrees of freedom, which corresponds to the non-precessing case. The lower the $p$-value, the more significant is the deviation from the non-precessing $\chi^2$-distribution.}
    \label{tab:rho_p}
\end{table*}

\clearpage

%%%%%%%%%%%%%%% BIBLIOGRAPHY
\bibliographystyle{apsrev4-1}
\bibliography{References,Populations}

\end{document}